\documentclass[preprint,3p]{elsarticle}

\let\today\relax
\makeatletter
\def\ps@pprintTitle{%
    \let\@oddhead\@empty
    \let\@evenhead\@empty
    \def\@oddfoot{\footnotesize\itshape
         {Submitted preprint} \hfill\today}%
    \let\@evenfoot\@oddfoot
    }
\makeatother

\usepackage{natbib}
\biboptions{authoryear}

\usepackage{epsfig}
\usepackage{amssymb}
\usepackage{amsmath}
\usepackage{psfrag}
\usepackage{units}

\usepackage[T1]{fontenc}
\usepackage[english]{babel}
\usepackage{ulem}
\usepackage{amsmath}
\usepackage{amsfonts}
\usepackage{amssymb}
\usepackage{graphicx}
\usepackage{subfigure}
\usepackage{color}
\usepackage[usenames,dvipsnames]{xcolor}

\usepackage{color}

\begin{document}

\begin{frontmatter}

\title{Damage spreading in quasi-brittle disordered solids: II. What the statistics of precursors teach us about compressive failure}

\author{Estelle Berthier}
\address{Institut Jean Le Rond d'Alembert (UMR 7190), CNRS and Sorbonne Universit\'e, 75005 Paris, France\\
Current address: Arnold-Sommerfeld-Center for Theoretical Physics and Center for NanoScience, Ludwig-Maximilians-Universit\"at M\"unchen, D-80333 M\"unchen, Germany}
\author{Ashwij Mayya and Laurent Ponson\corref{cor1}}
\ead{laurent.ponson@upmc.fr}
\address{Institut Jean Le Rond d'Alembert (UMR 7190), CNRS and Sorbonne Universit\'e, 75005 Paris, France}
\cortext[cor1]{Corresponding author}

\begin{abstract}
We investigate numerically and theoretically the precursory intermittent activity characterizing the preliminary phase of damage accumulation prior to failure of quasi-brittle solids.  We use a minimal but thermodynamically consistent model of damage growth and localization developed by \cite{Berthier_E}. 
The approach accounts for both microstructural disorder and non-local interactions and permits inferring 
a complete scaling description of the spatio-temporal structure of failure precursors. By developing a theoretical model of damage growth in disordered elasto-damageable specimen, we demonstrate that these scaling relations emerge from the physics of elastic manifolds driven in disordered media, while the divergence of these quantities close to failure is reminiscent of the loss of stability of the specimen at the localization threshold. Our study sorts out a long-standing debate on the nature of the compressive failure point and the origin of the universal statistics of the precursors preceding it. Our analysis rules out a critical-point scenario in which the divergence of the precursor size close to failure is signature of a second-order phase transition governed by the microstructural disorder. Instead, we show that while the jerky evolution of damage prior to failure results from the presence of material disorder, the latter does not significantly change the nature of the localization process, which is an instability well described by standard bifurcation theory of homogeneous systems. 
Finally, we harness our detailed understanding of the precursory statistics to design a methodology to estimate the residual lifetime of a structure from the statistical analysis of precursors. This method relevant for structural health monitoring is shown to perform rather accurately on our data. 
\end{abstract}

\begin{keyword}
Compressive failure \sep damage accumulation and localization \sep quasi-brittle materials \sep failure precursors \sep statistical aspects of failure \sep disordered materials \sep failure prediction \sep structural health monitoring
\end{keyword}

\end{frontmatter}

Quasi-brittle failure takes place through the accumulation and then the localization of a large number of microcracks in interaction. This failure mode is ubiquitous in a large range of materials such as concrete, rocks, woods, ceramics, especially under compressive loading conditions~\citep{Ashby,Lockner,Kachanov}. However, the way microcracks evolve and organize over time to ultimately lead to the failure of a specimen or a structure remains poorly understood.

Quasi-brittle failure generally proceeds in the following way~\citep{Fortin, Manzato, Tal}: First, damage spreads rather uniformly within the specimen. Then, microcracks progressively coalesce and organize into fracture patterns of increasing size. Finally, at the localization threshold, the damage activity concentrates into a macroscopic band, ultimately leading to the complete failure of the specimen. Noticeably, damage accumulation preceding localization is accompanied by an intense acoustic activity~\citep{Lockner2, Petri, Fortin2}. These acoustic emissions are reminiscent of bursts of damage activity, also referred to as {\it precursors}, separated by silent periods reminiscent of elastic reloading phases. As localization approaches, the intermittency in the damage growth, together with the acoustic activity intensify: The size and duration of the bursts both increase as a power law with the distance to failure~\citep{Guarino, Girard, Kun2, Baro2, Vu}. As a result, close to failure, they largely exceed the characteristic microstructural size of the material and its associated time scale. 

The statistical features of the mechanical and acoustic bursts have been extensively investigated both experimentally~\citep{Garcimartin, Guarino2, Deschanel, Davidsen, Rosti, Baro, Vu} and numerically~\citep{Roux2, Zapperi3, Tang, Amitrano4, Alava2, Pradhan3, Girard, Kun}. It turns out that they follow power-law statistics involving scaling exponents that are robust and independent, to a large extent, of the type of materials and loading conditions. These observations have fed a long-standing debate on the nature of failure by damage accumulation. Several studies interpreted failure as {\it a first-order transition}~\citep{Zapperi5, Alava2, daRocha}. While consistent with the observation of a loss of material rigidity at the failure point, this approach does not account for the universal scaling relations characterizing the precursors, like the power laws relating their size, duration and spatial extent, as well as the divergence of these quantities close to failure, two features that are generally associated with critical phenomena. As a result, it has also been argued that compressive failure can be interpreted as {\it a second-order phase transition} where the critical point corresponds to the failure threshold~\citep{Garcimartin, Moreno, Girard, Weiss2, Vu}. In this scenario, the scaling exponents involved in the precursors statistics are critical exponents. Their value depends on the range of the elastic interactions, while the divergence of the bursts size and duration close to failure is reminiscent of the classical phenomenology of critical phenomena. This scenario is particularly attractive as it entails universality of the scaling exponents. Yet, a theoretical proof of the critical nature of compressive failure in elasto-damageable solids is still lacking. As a result, the statistical features of the precursors observed experimentally, such as the value of the scaling exponents, still remain  largely unexplained.

 In the present work, we revisit this debate by performing a thorough analysis of the precursors statistics in a simplified model of failure by damage accumulation. We do not seek to provide a  comprehensive modeling of  the intermittent evolution of damage observed during compression experiments in all its complexity. Instead, we aim to identify and characterize the elementary mechanisms that underlie intermittency during compressive failure and account qualitatively for the statistical features of precursors observed experimentally. For that purpose,  we use a simplified damage model that qualitatively captures the various aspects of damage accumulation in disordered solids. We discuss the discrepancies between this simple model and real materials response and their implications in our interpretation.  We consider the simplified model of damage accumulation and failure proposed in~\cite{Berthier_E}, which we review in the first section. In this approach, quasi-brittle specimens are described at a continuum mesoscopic scale by a 1D array of interacting elasto-damageable elements loaded in parallel with randomly distributed damage thresholds. Interactions between neighboring elements emerging from elasticity are described qualitatively by introducing a cooperative length scale through a non-local damage variable~\citep{Pijaudier,Fremond,Pijaudier3}. This length scale governs the spatial extent of the stress redistribution taking place in the specimen after an individual damage event. As shown by~\cite{Berthier_E}, this approach is sufficiently simple to reveal and identify the basic mechanisms underlying quasi-brittle failure and rich enough to capture the main features of their mechanical response.  In particular, it predicts damage localization and catastrophic failure, which was showed to result from the unstable evolution of specific growth modes of the damage field that are selected by the spatial structure of the non-local interactions. Here, we first show that this model also provides a realistic description of the precursory activity preceding failure. We investigate the size, duration and spatial extent of damage bursts as predicted by this approach and demonstrate that they follow scaling relations. We also demonstrate that these quantities, on average, diverge as the specimen is driven closer to failure, thus reproducing qualitatively the main features observed experimentally~\citep{Garcimartin,Guarino2,Baro2,Vu}. 

Taking advantage of the simplicity of the considered model, we then explore theoretically the elementary mechanisms at the origin of the jerky evolution of damage during the accumulation phase preceding failure. We first retrieve that precursors are cascades of elementary failure events, also called {\it avalanches}, triggered by each other through the non-local stress redistribution following each individual event. We then derive an evolution equation of the damage field within the specimen and show that it behaves like an interface, the elasticity of which derives from the non-local nature of the interactions considered in our damage model. This interface is driven in a disordered medium reminiscent of the disordered fracture properties of the specimen with a driving speed that diverges as the specimen approaches failure. This mapping between quasi-brittle failure and the realm of driven elastic interfaces~\citep{Barabasi}, already theorized by~\cite{Weiss2} and~\cite{Vu}, successfully accounts for the statistical features of the precursors observed {\it at some finite distance} to failure. In particular, it explains the scaling relationships between the size, the duration and the spatial extent of precursors as observed in our simulations. However, at odds with ~\cite{Weiss2} and~\cite{Vu}, our analysis shows that the divergence of these quantities close to the localization threshold is not reminiscent of the depinning transition, a dynamic phase transition emerging from the competition between disorder and elasticity met by driven elastic interfaces~\citep{Narayan}. Instead, this behavior derives from the unstable nature of damage localization~\citep{Rudnicki, Bigoni, Dansereau}, a standard bifurcation that does not result from the presence of material disorder.

This interpretation of the divergence of the precursor size and duration close to failure has several important implications that are thoroughly discussed in our article. In particular, it implies that final failure does not result {\it per se} from the micro-instabilities observed during the precursory accumulation phase. Precursors are simply by-products of the collective growth of the damage field in interaction with the material disorder. As such, shutting down the disorder would also shut down the intermittency and would result in a smooth homogeneous growth of damage until the localization threshold is reached. From a modeling perspective, it means that the approach proposed by~\cite{Berthier_E} and extended more recently to 2D elasto-damageable solids by~\cite{Dansereau} using standard bifurcation theory of homogeneous systems is sufficient to account for damage localization and failure, as well as to predict the load bearing capacity of quasi-brittle solids with a reasonable accuracy. The proposed explanation of the divergence of the burst size and duration close to failure suggests that this mechanism is ubiquitous in compressive failure and follows the same law in a large range of materials. 

In the last part of our study, following the seminal ideas of~\cite{Sornette}, we harness this property to design a methodology that predicts the residual lifetime of structures from the statistical analysis of precursors. We test the approach on our numerical data set and show that the methodology successfully predicts failure from the evolution of the precursor sizes measured over a short loading window far from localization.  Hence, this approach paves the way for quantitative and predictive methods of structural health monitoring in more complex situations~\citep{Mayya}.

Our article is organized as follows. In the first section, we review the model proposed by~\cite{Berthier_E} which is used to simulate the intermittent evolution of damage in quasi-brittle solids. The second part is dedicated to the thorough statistical characterization of the precursors. Their magnitude, duration and spatial extent are related to each other by simple scaling laws and their variations over time display a power-law increase as failure approaches. The third section deals with the theoretical description of the intermittency and the interpretation of these properties. We provide an evolution law of the damage field that accounts for the collective evolution of the damage within the specimen and that fully captures the statistical properties observed numerically. This evolution equation sheds light on the complex connection between quasi-brittle failure and the theoretical framework of driven elastic interfaces. It also provides explanation for the divergence of the precursor size close to failure as a result of the unstable nature of the localization process. The last section is dedicated to the application of these concepts in structural health monitoring. We bring the numerical proof of concept that the non-stationary nature of the precursor statistics can be harnessed to predict the residual lifetime of progressively damaging specimens.

\section{Model and numerical implementation}\label{sec:Sec1}
In this section, we provide the main features of the model used numerically to investigate the intermittent evolution of damage during compressive failure. We consider a material description where each element damages  progressively owing to the microfracturing processes at lower length-scales. Hence, both heterogeneities in damage resistance and elastic redistributions after damage are described at a mesoscopic continuum scale. The reader is invited to refer to the work of~\cite{Berthier_E} for a more detailed review and discussion on the average failure behavior predicted by this approach. 

\subsection{A thermodynamics based damage model}\label{sec:Sec1a}

In our description, we consider a uni-dimensional structure ($\Sigma$) made of elasto-damageable elements in parallel. Such a system is loaded between two rigid plates: The bottom one is clamped while a quasi-static uni-axial loading, controlled in displacement $\Delta$, is applied to the top plate that experiences a reaction force $F$. We consider that each element is homogeneously deformed and undergoes a displacement equal to the macroscopic one. Individual elements are characterized by a scalar damage parameter $d$. This quantity, analogous to a microcracks density, ranges from zero when the element is intact to one when fully broken. Such a damage level affects the elastic response of each element as its elastic stiffness $k$ decreases with $d$. Each element represents a mesoscopic heterogeneity of randomly drawn damage energy $Y_c$ that vary with the damage level in the element. As a result, the total energy required to fully break an element is given by $\int_{0}^{1}Y_c(\tilde{d})d\tilde{d}$. The total energy of the system comprising the elasto-damageable specimen and the loading device is thus written as the sum of three contributions: The elastic energy $E^{el}$ stored in the materials elements, the energy $E^{d}$ dissipated due to damage growth and the work $W$ of the external force. It follows

\begin{equation}\label{eq:Etot}
E = E^{el}+E^d-W =  \int_\Sigma \dfrac{1}{2}\Delta^2k(\overline{d}(x))dx+\int_\Sigma\int_0^{d(x)}Y_c(x,\tilde{d})d\tilde{d}dx-\int_0^\Delta F(\tilde{\Delta})d\tilde{\Delta}.
\end{equation}

\noindent  As expressed in this equation, the local stiffness is chosen to depend on a non-local damage parameter $\overline{d}$ rather than on its local counterpart $d$. This non-local damage field corresponds to the weighed average 
\begin{equation}\label{eq:alpha}
\overline{d}(x) = \alpha(x) \ast d(x) = \int_\Sigma \alpha(x-x')d(x')dx'.
\end{equation}
As shown in~\cite{Berthier_E}, the introduction of such a non-local damage variable allows for a simple and practical implementation of the elastic interactions within the 1D structure. 
Indeed, each individual damage event triggers a redistribution of stress, the spatial extent of which is set by the so-called {\it interaction function} $\alpha(x)$. Interestingly, with this approach elastic interactions can be tuned to explore their impact on the failure response of the specimen, as done in ~\cite{Berthier_E}. In addition, this formulation ensures that the total energy is conserved all along the process of damage accumulation. 

A damage criterion is then obtained by differentiating Eq.(\ref{eq:Etot}) with respect to the damage parameters, and invoking energy conservation. In practice, the rate of mechanical energy composed of work of the external force and the elastic energy released in the specimen must compensate the  energy dissipated by damage for the damage increment. Denoting $\mathcal{F}(x)= -\dfrac{\delta E}{\delta d}$ the total driving force for damage, the damage criterion writes as

\begin{equation}
\label{eq:driving}
\left
\{ \begin{array}{lcl} \vspace{5pt}
\mathcal{F}(x) & = & \overline{Y}(x)-Y_c(x) < 0 \quad  \Rightarrow \quad \mathrm{No \, damage} \\
\mathcal{F}(x) & = & \overline{Y}(x)-Y_c(x) = 0 \quad \Rightarrow \quad \mathrm{Damage \, growth}
\end{array}
\right .
\end{equation}

\noindent where $\overline{Y}(x)$ is the non-local elastic energy release rate. It writes as the convolution of the interaction function $\alpha(x)$ with the local rate of energy restitution $Y(x)$ (see \cite{Berthier_E}):

\begin{equation}
\begin{split}
    \overline{Y}(x)  &= \alpha(x) \ast Y(x)~~\mathrm{where} \\
Y(x) &=  -\dfrac{1}{2}\Delta^2k'(d(x)).
\end{split}
\label{eq:Y}
\end{equation}
Here, $k'$ denotes the derivative of the stiffness with respect to the damage parameter $d$. Note that we adopt here a thermodynamically-consistent damage mechanics framework~\citep{Fremond,Pham}, in a similar manner to fracture mechanics predicting crack propagation from the balance of mechanical and fracture energy~\citep{Lawn,Rice8}. The particularity of the proposed model relies in that the driving force involved  in the damage criterion is non-local. This arises from the  dependency of the local stiffness on the non-local damage parameter. It means that the driving force in one element depends on the damage level  within the surrounding elements, in a manner defined by the interaction function $\alpha$. It also implies that an increase of damage in one element induces an update of the damage driving force in an extended region around the damaging element. Therefore, the tunable function $\alpha$ controls the mechanism of stress redistribution following damage events.  

Energy conservation is illustrated in Fig.~\ref{fig:Fig1}(a). Below the elastic limit $\Delta_{\mathrm{el}}$, the elastic energy release rate is not sufficient to damage the material: therefore the work of the external force is entirely stored as elastic energy. Above $\Delta_{\mathrm{el}}$, the mechanical energy injected within the medium via the work of the external force is shared among stored elastic energy and released dissipated energy due to damage. As we will see in the following, the dissipation of mechanical energy through damage occurs through sudden bursts separated by silent elastic phases. This intermittency during quasi-brittle failure of disordered materials is the main focus of this study.

\subsection{Material parameters}
The microstructural heterogeneities of the material are accounted for by spatial variations in the damage energy field $Y_c(x,d)$, described by the quenched noise $y_{\mathrm{c}}(x,d)$. For each element located at $x$ and having a damage level $d$, $y_{\mathrm{c}}(x,d)$ is drawn from a uniform distribution of standard deviation $\sigma$ and zero mean value. The field of resistance to damage hence writes as
\begin{equation}\label{eq:Yc}
Y_c(x,d) = Y_{c0}\left[1+y_\mathrm{c}(x,d)+\eta \, d\right]
\end{equation}
    \noindent where $\eta>0$ is a hardening parameter and $Y_{c0}$ is the average damage energy of the intact material. Such a toughening behavior has been observed experimentally~\citep{Berthier_E3,Renard2} and considered in other damage models~\citep{Pham3}. This hardening behavior may be related to our modeling choice that consists in describing damage growth at a mesoscopic scale. Indeed, as the mesoscopic element progressively damages, the distribution of damage thresholds within this material element is explored from its lowest values to its highest ones. As a result, the value of the parameter $\eta$ may emerge from the spreading of microcracking thresholds at the microscopic scale.
    
    In terms of statistical modeling, our description can be seen as intermediary between original fiber bundle models comprising brittle fibers and continuum damage mechanics model with homogeneous resistance to failure. It allows exploring material behaviors with extended post-peak response.\footnote{ Small hardening parameters $\eta \simeq 0$ result in a brittle failure of the material element and, as a consequence, of an abrupt failure of the specimen after the first elements break down The population of precursors in this case is dominated by the material disorder: Catastrophic and sudden failure without precursors is observed for weakly disordered solids whereas strong disorder leads to delayed fracture with power-law distributed precursors~\citep{Shekhawat}.}  The mechanical response remains qualitatively unaffected for a broad range of parameter values as long as disorder is present and a transition from a stable to an unstable growth of damage in the constituting elements takes place as $d(x)$ increases (see ~\cite{Berthier_E} for the determination of the ranges of material parameter verifying this condition). 

The stiffness decay with damage, representative of the degradation of the elastic properties as microcracks develop, is introduced via a polynomial expression. For a single element, such a softening writes as 
\begin{equation}\label{eq:k}
k(d) = k_0\left[ad^3-(a+1)d+1\right]
\end{equation}
\noindent where $k_0$ is the stiffness of the intact material and $a$ is a constant the sign of which, in combination with $\eta$, sets the stability of individual elements (see \cite{Berthier_E}). 
Such a polynomial expression allows for the exploration of a large range of mechanical response at the element scale. Different softening laws $k(d)$ can arise from a large variety of microscopic damage processes, and we can for example refer to~\cite{Kachanov4} for a thorough investigation of the dependency of the elastic moduli on the considered damage processes. Moreover, in combination with an appropriate choice of hardening parameter $\eta$, the expression~\eqref{eq:k} with $a < 0$ may result in an extended regime of stable damage growth even in the limit of small disorder, unlike in conventional fiber bundle models where such a ductile-like response is strongly dependent on the disorder level~\citep{Shekhawat,daRocha}. The linear decrease of the elastic modulus with damage (recovered here by setting $a=0$) classically used in continuum damage mechanics~\citep{Lemaitre} leads to {\it local} damage models that do not capture the intermittent evolution of damage in disordered materials. Nonlinear expressions as used in~\cite{Girard} and~\cite{Thilakarathna} results in non-local interactions during damage growth, giving rise to more realistic dynamics without losing the generality of the model. 

The interaction function $\alpha$ introduced in Eq.~\eqref{eq:alpha} to define the non-local damage parameter is expressed as
\begin{equation}\label{eq:alpha}
\alpha(x) = \alpha_0 \exp\left(-\dfrac{\vert  x \vert}{2\ell}\right)
\end{equation}

\noindent In this expression, $\vert x \vert$ represents the distance between elements, $\alpha_0$ is a normalization constant ensuring energy conservation such that$\int_\Sigma \alpha(x)dx = 1$ and $\ell_o$ is an internal length which controls the spatial extent of interactions within the specimen. In particular, it provides the range over which the driving force is redistributed after a damage event. Here also, as illustrated subsequently, the main results of our study remain robust to the choice of $\ell_o$ as long as it is much larger than the element size and much smaller than the specimen size. In practice, we consider internal lengths several times larger than the heterogeneity size and system sizes several thousand times larger. While real materials display long-range interactions~\citep{Bazant9, Dansereau}, similar forms of interaction functions have been assumed in non-local models of damage evolution in quasi-brittle materials~\citep{Fremond,Pham, Pijaudier3}. The expression in Eq.~(\ref{eq:alpha}) permits analytical calculations of failure and localisation threshold and  qualitatively captures the main aspects of damage evolution resulting from the cooperative action of disorder and interactions. 


\subsection{Numerical resolution}\label{sec:numerical}
We take advantage of the quasi-static loading conditions and adopt an extremal dynamics inspired by \cite{Schmittbuhl4} based on the following rules: The imposed displacement is increased until the failure criterion is reached for at least one element. The damage level of this element is then increased by an elementary increment $\delta d_0 \ll 1$, and the spatial distribution of driving forces in the material is updated using Eqs.~(\ref{eq:Y}), (\ref{eq:Yc}) and (\ref{eq:alpha}). As schematically shown in  Fig.~\ref{fig:Fig1}(d), the redistribution of driving force results in additional damage, leading to  additional driving force redistributions until such a cascade stops. This occurs when the driving force $Y$ is below its critical value $Y_c$, everywhere in the specimen. The external load is then further increased until the weakest element of the array reaches its failure threshold, triggering another cascade of damage events.

 To simulate damage growth, we consider a discretized specimen made of $N$ elasto-damageable elements loaded in parallel. Initially, damage $d(x) = 0$ is zero everywhere and the field $Y_c(x,d)$ of damage resistance is randomly distributed owing to the disorder term $y_c(x,d)$. We  compute the non-local damage field described in Eq.~\eqref{eq:alpha} using the convolution property of Fourier transforms, 

\begin{equation}
    \overline{d}(x) = \mathrm{F}^{-1}(\tilde{\alpha}(q) \times \tilde{d}(q))
\end{equation}
\noindent where, $\mathrm{F}^{-1}$ denotes the inverse Fourier transform and $\tilde{\alpha}$ is the Fourier transform of the function $\alpha$. This is then used to calculate the damage driving force of Eq.~\eqref{eq:Y}
\begin{equation}
    \overline{Y}(x) = -\frac{1}{2}\mathrm{F}^{-1}(\tilde{\alpha}(q) \times\tilde{k'}(\overline{d}) )\Delta^2 .
\end{equation}

\noindent The displacement $\Delta$ is increased such that exactly one  element satisfies the damage criterion $\overline{Y}(x,d) = Y_c(x,d)$. The damage level for this element is then increased by $\delta d_0 \ll 1$. After that, the field of damage driving force is recalculated. The heterogeneous field of damage energy is also redrawn after each damage event as
\begin{equation}
    Y_c(x,n_0) = Y_{c0}[1 + y_c(x,n_0) + n_0 \overline{\eta}]
\end{equation}

\noindent where, $n_0 = d/\delta d_0$ is the total number of instances that an element has damaged, $\overline{\eta} = \eta\delta d_0$ and $y_c(x,n_0)$ is a random number drawn from a centered uniform distribution with standard deviation $\sigma$. Additional damage events may follow due to the redistribution. The cascade process stops when $\overline{Y}(x) < Y_c(x,d)$ for all elements. Macroscopically, a load drop at constant displacement is registered due to the increase of the damage level. The displacement is then again increased so that one element satisfies the damage criterion. This sequence of damage growth and reloading continues till damage variable is one in all material elements. 

There is no explicit time in the simulation. Yet, we can define \textit{a posteriori} the rate $v_{\mathrm{ext}}$ at which the displacement imposed to the specimen is increased. We consider the limit $v_{\mathrm{ext}} \rightarrow 0$ that corresponds to quasi-static loading conditions. Thus, the elastic reloading phases are much longer than the dissipative phases. This justifies our numerical procedure. In addition, it results in two specific features: Damage cascades take place under fixed imposed displacement and precursors are well separated in time so they appear as isolated dissipative events. In the following, we use $\eta = 9$, $a = -0.3$, $\delta d_0 = 0.005$ and $\sigma = 0.2$. The reader is invited to refer to \cite{Berthier_E} for a detailed study of the effect of the parameter values $\eta$ and $a$ on the damage evolution. Note however that the statistical properties of the precursors revealed by our study remain identical for other parameter values as long as failure is preceded by a preliminary regime of damage accumulation.

\section{Intermittency and precursors statistics}\label{sec:Sec2}

\subsection{Global and local scale manifestations of the intermittent growth of damage}

\begin{figure}
\center
\includegraphics[width = .85\textwidth]{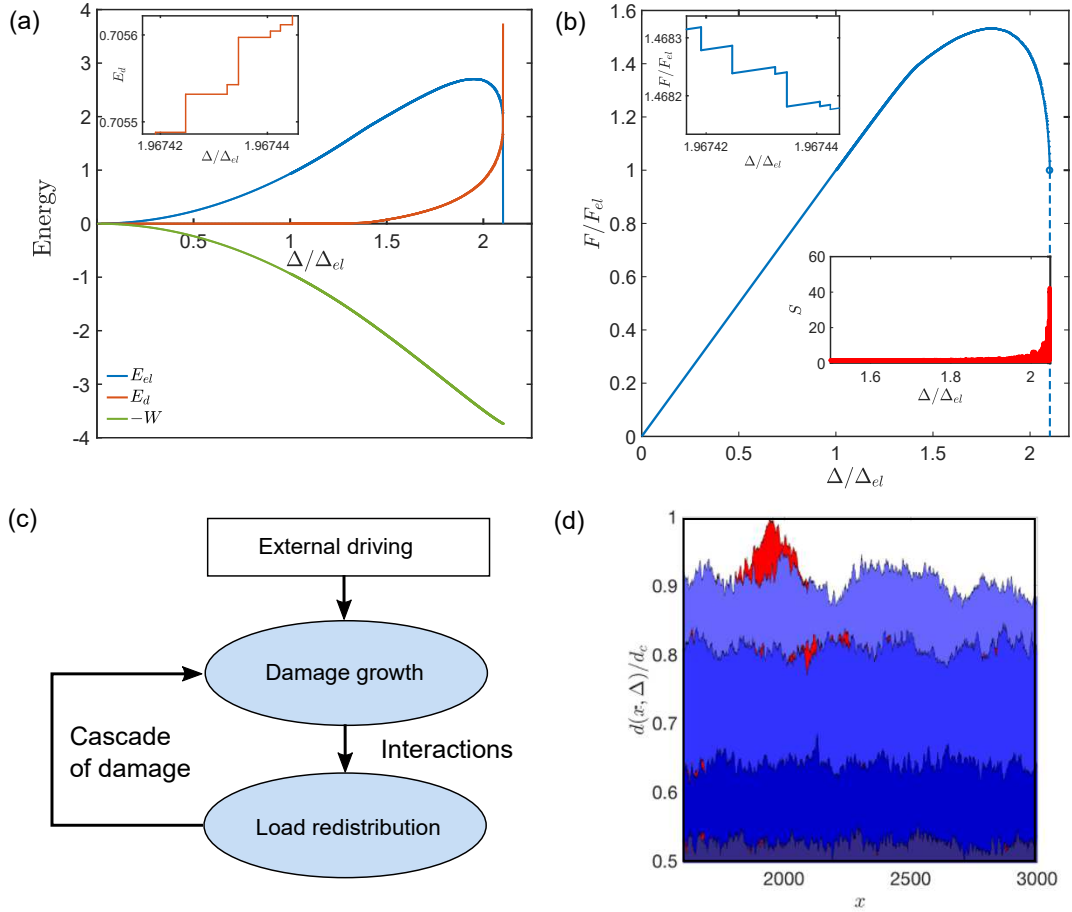} 
\caption{Observation of precursors and basic damage mechanism: (a) Evolution of the elastic energy $E_\mathrm{el}$, the dissipated energy $E_\mathrm{d}$ and the work $W$ of the external force with the imposed loading amplitude $\Delta$, normalized by the elastic limit $\Delta_\mathrm{el} = \sqrt{-2Y_{c0}/k'(0)}$. Note that the sum of the three terms remains constant, as a result of the energy conservation principle from which derives our model. The inset emphasizes the intermittent nature of damage evolution that displays bursts of dissipated energy; (b) macroscopic force-displacement response during damage growth normalized by the elastic limits $F_{el}$ and $\Delta_\mathrm{el}$ of the specimen displaying catastrophic failure for $\Delta_\mathrm{c} \approx 2.1 \, \Delta_\mathrm{el}$. The size $S$ of damage bursts is shown in the lower inset as a function of the distance to failure. The effect of the damage cascades on the force-displacement response of the specimen is evidenced by the sudden force drops shown in the upper inset; (c) schematic of the feedback loop at the origin of the precursory micro-instabilities; (d) damage field before (blue) and after (red) ten successive avalanches at different distances to failure. The damage field is normalized by the critical damage level $d_\mathrm{c}$ at failure.}
\label{fig:Fig1}
\end{figure}

As observed experimentally by~\cite{Petri}, \cite{Guarino}, \cite{Davidsen}, \cite{Baro} or \cite{Vu}, the energy dissipated through damage during quasi-brittle failure evolves in a step-wise fashion, even though the applied load amplitude is continuously increased. This phenomenon is captured by our approach, as illustrated in the inset of Fig.~\ref{fig:Fig1}(a). The mechanical response of the specimen under slowly increasing load consists of slow reloading phases where the material behaves elastically (identified by plateau regimes in the evolution of the accumulated dissipated energy) and micro-instabilities taking place at constant displacement (identified by instantaneous increase of the accumulated  dissipated energy). Owing to the description of damage as a transfer of mechanical energy into dissipated energy, the evolution of the elastic energy stored in the specimen also reflects such a jerky dynamics (not shown here). The signature of this intermittent behavior is also observable from the macroscopic force-displacement curve, as shown in the inset of Fig.~\ref{fig:Fig1}(b). Here, the elastic reloading phases are separated by sudden drops of force resulting from the decrease in elastic stiffness associated with the degradation of the material. As further highlighted later, a remarkable feature of these fluctuations is the broad range of scales that they cover. This is in contrast with standard Gaussian statistics that describes mild fluctuations, for example observed  in at-equilibrium systems, and that are reminiscent of thermal fluctuations.

Such an intermittent dynamics, also referred to as {\it crackling noise}~\citep{Sethna}, is reported for a large range of mechanics problems involving disorder, such as tensile failure of brittle disordered solids through crack growth~\citep{Ponson19}, wetting of disordered surfaces~\citep{Eggers} or imbibing of porous media~\citep{Planet}. It is also observed beyond mechanics, for example during the magnetization of ferromagnets, displaying the so-called Barkhausen noise~\citep{Zapperi2}, or even in social systems as illustrated by the application of these concepts for describing the abnormally large fluctuations of the stock markets~\citep{JPBouchaud2}. As a common denominator to all these very different physical systems, long-range interactions in interaction with disorder result in feedback loops that trigger micro-instabilities with broadly distributed sizes.

The mechanism underlying such {\it avalanches} in compressive failure is illustrated in Fig.~\ref{fig:Fig1}(c). It consists of a cascade of local failure events triggered by each other by the redistribution of stress in the specimen that follows each elementary damage. A remarkable feature of these cascades is the evolution of their energy (referred to as their size $S$) that increases as the specimen approaches failure, as shown in the inset of Fig.~\ref{fig:Fig1}(b). This evolution reveals the progressive loss of stability of the specimen, a connection further explored in Section~\ref{sec:Sec3}.

Beyond the time-series features of the damage evolution, we also investigate the spatial structure of the damage cascades. To that end, we study damage spatial organization at the local scale by comparing the damage field before (blue) and after (red) a certain number (typically ten) of successive avalanches, as illustrated in Fig.~\ref{fig:Fig1}(d) for four different load levels. Far from failure, {\it i.e.} at low damage levels $\langle d\rangle \lesssim 0.55$, damage spreading is dominated by material disorder. Damage grows rather uniformly within the specimen, and the incremental variation of the damage field is characterized by small clusters of dimension comparable with the characteristic size of the material heterogeneity. On the contrary, close to failure for $\langle d\rangle \simeq 0.9$, we observe compact localized avalanches, the size of which is much larger than the heterogeneity size. At this stage, the redistribution process dominates over material disorder, leading to large and structured precursors. This competition between material disorder and elastic interactions, as well as its evolution as the specimen is driven towards failure, will be thoroughly investigated in the following. 

\subsection{Statistical characterization of precursors}\label{sec:Statistics}
\begin{figure}[h!]
\center
\includegraphics[width = \textwidth]{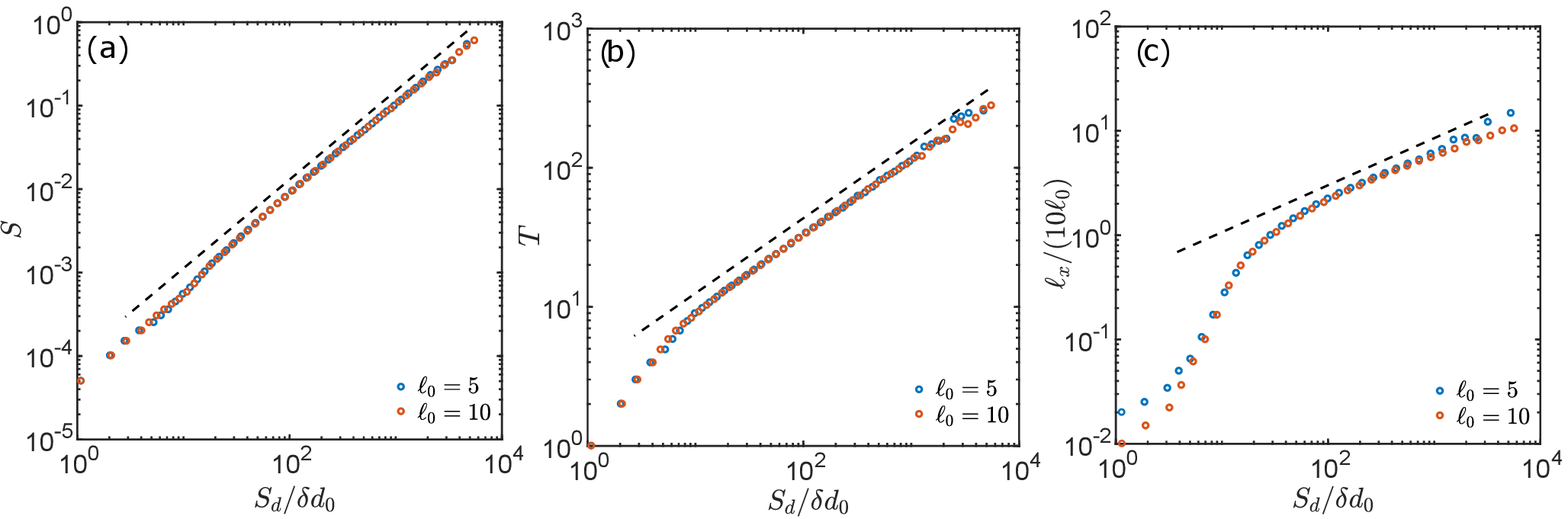} 
\caption{Scaling behavior of precursors as observed during the whole process of damage evolution: Relationship between (a) the dissipated energy  $S$, (b) the duration $T$  and (c) the spatial extent $\ell_\mathrm{x}$ of the precursors to the total number $S_\mathrm{d}$ of elementary damage increments per cascade. 
The dashed lines indicate the scaling relationships $S \sim S_\mathrm{d}$, $T \sim S_\mathrm{d}^{z/d_\mathrm{f}}$ and $\ell_\mathrm{x} \sim S_\mathrm{d}^{1/d_\mathrm{f}}$, respectively, leading to the value of the fractal dimension $d_\mathrm{f} = 2.35 \pm 0.15$ and the dynamic exponent $z = 1.40 \pm 0.15$.}
\label{fig:Fig2}
\end{figure}

To investigate the intermittent evolution of damage, we characterize each precursor by three quantities: its spatial extent $\ell_\mathrm{x}$, its size $S_\mathrm{d}$ and its duration $T$. $\ell_\mathrm{x}$ corresponds to the distance between the two most distant damaged elements belonging to the same damage cascade. $S_\mathrm{d}$ is defined as the total number of elementary damage increments (the quantity $\delta d_0$ in Section ~\ref{sec:numerical}) constituting the cascade. As shown from its linear variation with the dissipated energy $S$ during the whole cascade (see Fig.~\ref{fig:Fig2}(a)), $S_d$ measures the energy dissipated through a precursor. In the following, we consider equivalently $S$ or $S_\mathrm{d}$ to measure the precursor size. Despite the discrete dynamics adopted in our numerical scheme, we can also define the duration $T$ of a precursor from the total number of damage redistribution loops involved during the cascade. This amounts to assume that an individual damage event and the resulting redistribution of driving force take place over some material specific time scale that is much smaller than the characteristic time of the applied driving, an assumption that is supported by the limit $v_\mathrm{ext} \rightarrow 0$ considered in this study.

Remarkably, these three quantities are related through power laws as shown in Fig.~\ref{fig:Fig2}. In particular, the spatial extent $\ell_\mathrm{x}$ of the precursors is related to its size $S_\mathrm{d}$ via the fractal dimension $d_\mathrm{f}$ as
\begin{equation}\label{eq:ellx_vs_S}
\ell_x\sim S_\mathrm{d}^{1/d_\mathrm{f}}
\end{equation}
\noindent where $d_\mathrm{f} = 2.35 \pm 0.15$. As shown by the renormalization of the ordinates of Fig.~\ref{fig:Fig2}(c), $\ell_\mathrm{x}$ scales linearly with the interaction length $\ell_\mathrm{0}$ introduced in the redistribution function. Note also that the smallest precursors following such a scaling behavior are of the order of $10 \, \ell_\mathrm{0}$, suggesting that such damage cascades emerge from the cooperative response of a sufficiently large number of material elements in interaction.
Investigating now the duration of the precursors in Fig.~\ref{fig:Fig2}(b), we observe that it scales with the avalanche size as
\begin{equation}\label{eq:T_vs_S}
T\sim S_\mathrm{d}^{z/d_\mathrm{f}}
\end{equation}
\noindent where $z = 1.40 \pm 0.15$ is the so-called dynamic exponent. Therefore, these three quantities (size, duration and length scale) can be related to each other and, by virtue of these scaling, we can equally use one or the other to characterize the magnitude of a damage cascade.

As noticed previously, the intermittency during damage spreading is non-stationary. We now explore this feature through the detailed analysis of the avalanche size distribution computed at different distances $\delta$ to catastrophic failure. We define $\delta$ as 
\begin{equation}\label{eq:delta}
\delta = \dfrac{\Delta_\mathrm{c}-\Delta}{\Delta_\mathrm{c}-\Delta_\mathrm{el}}
\end{equation}
where $\Delta_\mathrm{el}$ and $\Delta_\mathrm{c}$ are the elastic limit and the failure load, respectively. This quantity ranges from one as the first damage event takes place to zero at failure.

\begin{figure}
\center
\includegraphics[width = .9\textwidth]{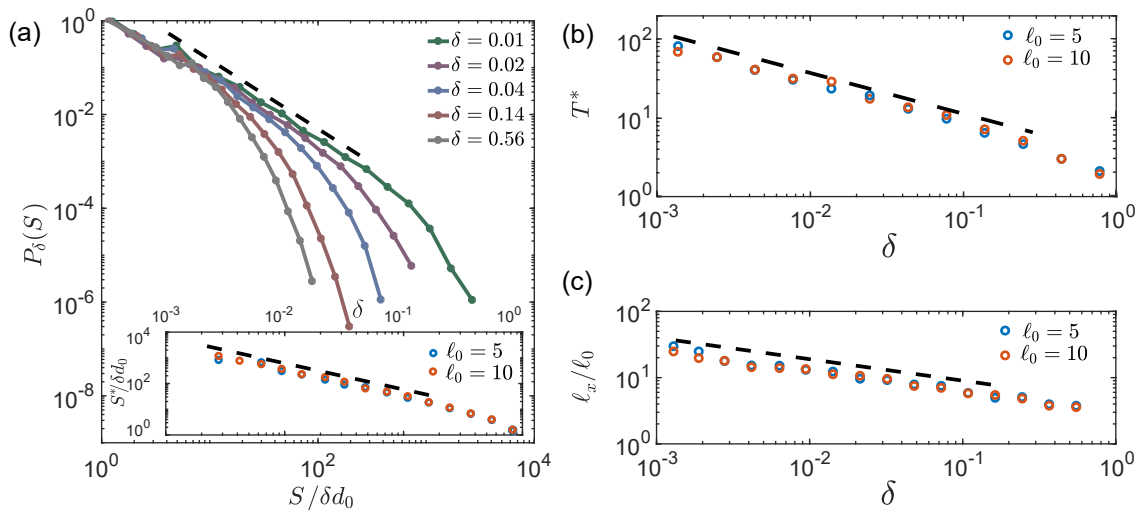} 
\caption{Statistics of avalanches at different distances to failure: (a) normalized probability density function computed at different distances $\delta$ to catastrophic failure. The dashed line indicates the scaling behavior $S^{-1.5}$. The variations of the cutoff avalanche size $S^*$ as defined by Eq.~\eqref{eq:PS_S} is shown in inset. It varies as $S^\star \sim 1/\delta$ as highlighted by the dashed line; (b) evolution of the cutoffs duration $T^\star$ and spatial extent $\ell_x^\star$ extracted from the distributions $P_\delta(T)$ and $P_\delta(\ell_\mathrm{x})$ (not shown here). They also increase as a power-law $T^\star\sim 1/\delta^{\phi}$ and $\ell_\mathrm{x}^\star \sim 1/\delta^{\kappa}$ with the distance to failure with the exponents $\phi \simeq 0.53$ and $\kappa \simeq 0.37$.}
\label{fig:Fig3}
\end{figure}

The precursors statistics is then investigated at different instants during the progressive failure of the specimen. To do so, we divide the specimen's lifetime into bins and, considering several realizations of the disorder, we compute the distribution of precursor sizes in each bin. Because the focus of this paper is on the precursors to failure and not the  failure event itself, the final cascade during which catastrophic failure takes place is left aside. 
The distributions $P_\delta(S)$ of avalanche sizes are shown in Fig.~\ref{fig:Fig3}(a) for $\ell_0 = 5$. Irrespective of the value of the interaction length, a power law distribution with exponent $\beta = 1.5\pm 0.1$ is obtained for avalanches of size $S \ll S^\star$. Above the cut-off $S^\star$ which increases as failure is approached, the probability density decays exponentially fast. Hence, the avalanche size distributions are well described by
\begin{equation}\label{eq:PS_S}
P_\delta(S) \sim S^{-\beta}e^{-S/S^\star}.
\end{equation}
 
The variations of $S^*$ with the distance to failure $\delta$ is shown for the interaction lengths $\ell_0 = 5$ and $\ell_0 = 10$ in the inset of Fig.~\ref{fig:Fig3}(a). In both cases, a power-law behavior
\begin{equation}\label{eq:S_vs_delta}
S^*\sim 1/\delta^{\gamma}
\end{equation}
\noindent is observed with $\gamma = 1.0 \pm 0.1$.

This property rationalizes the observation made in the inset of Fig.~\ref{fig:Fig1}(b): As the specimen approaches failure, the characteristic precursor size set by the cut-off $S^\star$ increases. Owing to the scaling relations shown in Fig.~\ref{fig:Fig2}, it implies that the characteristic spatial extent and characteristic duration of the precursors also increase as failure is approached. This is indeed verified in Fig.~\ref{fig:Fig3}(b): The cutoffs $T^\star$ and $\ell^\star$ as extracted from the distributions of duration $P_\delta(T)$ and spatial extent $P_\delta(\ell_\mathrm{x})$ of avalanches also increase with the distance to failure as
\begin{equation}\label{eq:ellx_vs_delta}
\left \{
\begin{array}{lcl} \vspace{3pt}
T^\ast & \sim & 1/\delta^{\phi} \\
\ell_\mathrm{x}^\ast & \sim & 1/\delta^{\kappa}
\end{array} \right .
\end{equation}
\noindent with $\phi = 0.53 \pm 0.10$ and $\kappa = 0.37 \pm 0.10$.

We can now connect these relationships with the ones previously evidenced in Fig.~\ref{fig:Fig2} between the size, the duration and the spatial extent of {\it all} the precursors. Applying them to the cut-off of the statistical distributions at some given distance $\delta$ to failure, one obtains $T^\star \sim (S^\star)^{z/d_\mathrm{f}}$ and $\ell_\mathrm{x}^\star \sim (S^\star)^{1/d_\mathrm{f}}$ that, together with Eqs.~\eqref{eq:S_vs_delta} and~\eqref{eq:ellx_vs_delta}, lead to $T^*(\delta)\sim  (S^*(\delta))^{z/d_\mathrm{f}}\sim\delta^{-\gamma z/d_\mathrm{f}}$ and $\ell_x^*(\delta)\sim (S^*(\delta))^{1/d_\mathrm{f}}\sim\delta^{-\gamma/d_\mathrm{f}}$ from which we infer the relationships between exponents
\begin{equation}\label{eq:T_vs_delta}
\left \{
\begin{array}{lcl} \vspace{3pt}
\phi & = & \gamma \, z/d_\mathrm{f} \\
\kappa & = & \gamma/d_\mathrm{f}.
\end{array} \right .
\end{equation}
Interestingly, a third relationship $\phi = \kappa \, z$ relating the exponents $\phi$ and $\kappa$ characterizing the non-stationary nature of the damage evolution with the dynamic exponent $z$ can be derived from the previous equations. We can verify that all these scaling relations are consistent with the values of the exponents measured in our simulations.

We now focus on the spatial characterization of the precursors at the local scale. As the specimen is driven closer to failure, damage cascades extend over material regions of increasing size. We aim to find the footprint of this increasing cooperative length scale on the accumulated damage field by introducing the \textit{correlation function}
\begin{equation}\label{eq:FonctionCorrelation}
C(\delta x) = \sqrt{\langle\langle d(x)\rangle^2_{\delta x}\rangle_x-\langle\langle d(x)\rangle_{\delta x}\rangle^2_x} = \mathrm{std}({\langle d(x) \rangle_\mathrm{\delta x}}).
\end{equation}

This function is then averaged over different disorder realizations, at some given distance $\delta$ to failure. From the central limit theorem, one expects the correlation function to decay as $1/\sqrt{\delta x}$ for an uncorrelated damage field, as awaited from a purely random process of damage growth. On the contrary, if the damage field is correlated over some length scale, a deviation from this behavior is expected. Hence, this correlation function reveals deviation from pure random processes and determines the scales at which such deviations take place.

\begin{figure}[ht!]
\center
\includegraphics[width = .55\textwidth]{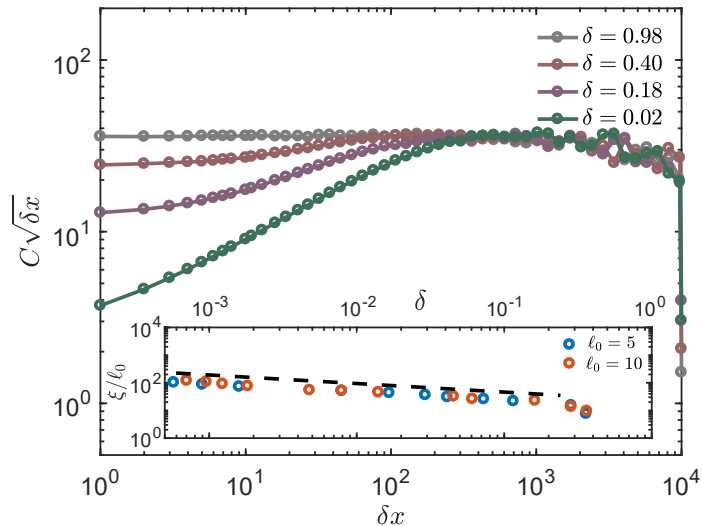} 
\caption{Normalized correlation function of the cumulative damage field at several distances $\delta$ to failure. The inset shows the evolution the extracted correlation length $\xi$ that is found to vary as $\xi \sim 1/\delta^{\rho}$ with $\rho \simeq 0.35$.}
\label{fig:Fig4}
\end{figure}

Figure~\ref{fig:Fig4} shows the correlation function as a function of the box size $\delta x$ after normalization by $\sqrt{\delta x}$, for $\ell_0 = 5$ and various distances to failure. As expected, far from failure (see for example $\delta = 0.98$), the normalized correlation function is constant, pointing out an uncorrelated damage field. As failure is approached, a deviation from this plateau behavior is observed. The crossover length $\xi$ between a square root behavior $C(\delta x) \sim 1/\sqrt{\delta x}$ at large length scales ($\delta x > \xi$), and a non-trivial regime at shorter length scales ($\delta x < \xi$), defines the correlation length $\xi$ of the damage field. At a given distance to failure, the cross-over is defined as the length-scale at which deviation from the plateau behavior is larger than $5\%$. The cut-off was not clear for early stages of damage and therefore, not considered for studying the variation of correlation length with distance to failure. The spatial extent of correlated damage increases as failure is approached, and so does the amplitude of the deviation from the random field behavior. Hence, the accumulated damage field shows correlations over an increasing range of length scales, up to the correlation length $\xi$. The latter increases as failure is approached and follows a power law
\begin{equation}\label{eq:xi_vs_delta}
\xi \sim 1/\delta ^\rho
\end{equation}

\noindent where $\rho =  0.35 \pm 0.10$ (see the inset of Fig.~\ref{fig:Fig4}). Interestingly, this behavior is similar to the one of the avalanche spatial extent that also increases as $\ell_x^*\sim 1/\delta^{\kappa}$ with $\kappa \simeq \rho$, suggesting that both quantities are signatures of the same process (see Fig.~\ref{fig:Fig3}(c)). Moreover, the lowest value of $\xi$ that could be identified at the earliest stage of the damage spreading is of the order of $10 \, \ell_0$, a length scale that compares with the precursor spatial extent above which scaling behaviors between size, duration and spatial extent starts to emerge (see Figs.~\ref{fig:Fig2}(c) and~\ref{fig:Fig3}(b)). Hence, we conclude that the cooperative dynamics of damage, which accumulates through bursts localized over some material region of characteristic size $\xi$, shapes the accumulated damage field through the introduction of non-trivial correlations over a length scale $\ell_\mathrm{x} \simeq \xi$ that increases as failure is approached.

In summary, the failure precursors can be characterized by the dissipated energy per avalanche $S$, the accumulated damage $S_d$, the duration $T$ or the spatial extent $\ell_x$ that are related all together by  scaling relationships. The scaling exponents involved are independent of the spatial extent of the redistribution function, in the range investigated in our study. In addition, the statistical characterization of precursors reveal that the progressive degradation of the specimen is a non-stationary phenomenon. In particular, the characteristic size, duration and spatial extent of the precursors diverges as a power law with the distance to catastrophic failure. The presence of large structured avalanches is encoded in the accumulated damage field from which emerges also a diverging correlation length Eq.~\eqref{eq:xi_vs_delta}. 

Overall, our detailed analysis of the precursors statistics reveals the existence of a time scale $T^\star$ and a length scale $\ell_\mathrm{x}^* \simeq \xi$ encoded in the damage field fluctuations. Both  characterize the process of damage spreading and diverge at failure. Such a behavior is often considered as a hallmark of critical phenomena. In this scenario proposed by~\cite{Garcimartin}, \cite{Moreno}, \cite{Girard}, \cite{Weiss2} and \cite{Vu}, compressive failure is interpreted as a second-order phase transition between a specimen able to sustain a mechanical load and a fully broken specimen. This interpretation is appealing as the observation of universal scaling exponents is a key feature of second order phase transitions. Nevertheless, and despite the strong similarities with the phenomenology of critical phenomena, we will see in the following that this conclusion is too hasty and that the behavior of the precursors as the specimen is driven closer to failure is actually not consistent with the response of a system that is driven towards a critical point. 


\section{Theoretical modeling : Deciphering the statistics of precursors}\label{sec:Sec3}
We now aim at providing a quantitative explanation for the statistical properties of the precursors observed in our simulations.
We first rewrite the damage model described in Section 1.1 under the form of an evolution equation of the damage field at a fixed distance to failure. This evolution equation describes strictly the same damage process and can be numerically resolved following the procedure described in Sec. 1.3. Yet, this evolution equation, which includes all the model ingredients described in Section 1, reveals {\it explicitly} the connection between the theoretical framework of driven disordered elastic interfaces \citep{Narayan, Barabasi, Leschhorn2} and the process of damage growth in disordered elasto-damageable solids, a connection already theorized by~\cite{Weiss2}. 
First, this mapping is used in Section~\ref{sec:Sec3b} to interpret the statistical features of the precursors observed in our simulations at a {\it fixed} distance to failure, such as the scaling relations between the size, the duration and the spatial extent of the damage cascades. In a second step, in Section~\ref{sec:Sec3c}, we explicitly write the dependance of each term of the evolution equation with the distance to failure and analyse it to determine the origin of the non-stationary evolution of precursors statistics observed in our simulations in Section~\ref{sec:Statistics}. 
We show that some parameters controlling the damage evolution equation, for example the driving speed of the elastic interface, diverge at failure. This observation accounts for the divergence of the duration and spatial extent of the precursors observed in our simulations, without invoking the role of material disorder. Our analysis thus rules out a scenario in which quasi-brittle failure is interpreted as a critical phenomenon emerging from the competition between elasticity and disorder and provides an alternative explanation for the divergence of the characteristic time and length scale of the damage cascades close to failure.

\subsection{Evolution equation of the damage field under fixed loading amplitude}~\label{sec:Sec3a}
We start our theoretical analysis by deriving an evolution equation of the damage field as it evolves from a reference damage level $d(x,t_0)=d_0$ at the reference time $t_0$ under a {\it fixed} loading amplitude $\Delta = \Delta_{\mathrm{0}}$. We consider the damage field perturbation $\delta d(x,t) = d(x,t) - d_0$ and assume a so-called over-damped dynamics $\dot{d}(x,t) = \dot{\delta d}(x,t) \propto \mathcal{F}(d(x,t),\Delta_0)$ where $\mathcal{F}$ is the damage driving force introduced in Eq.~\eqref{eq:driving}. This  assumption is commonly used for deriving kinetic laws from thermodynamic driving force, {\it e.g.} in brittle fracture or plasticity problems to relate the crack speed~\citep{Gao,Ponson20} or the plastic flow~\citep{Puglisi} to the local driving. As shown by~\cite{Chopin5} in the context of crack propagation under tensile loading conditions, it actually derives from the linearization of the rate-dependency of the dissipative term. This amounts here to assume that the resistance to damage $Y_\mathrm{c}$ is an increasing function of the damage rate $\dot{d}$.

We then decompose the total damage driving force defined in Eq.~\eqref{eq:driving} into two contributions: (i) A homogeneous term $\mathcal{F}_{hom} = Y(d_0,\Delta_0) - Y_\mathrm{c}(d_0)$ (where $Y$ and $Y_c$ are respectively given by Eqs.~(\ref{eq:Y}) and~(\ref{eq:Yc}) ), that depends only on the imposed loading amplitude (or equivalently to the distance to failure) and the initial damage level $d_0$, and (ii) an inhomogeneous contribution $\delta\mathcal{F}$ that depends on the damage field  perturbations $\delta d(x)$. The perturbation is assumed small in comparison to the average damage level. This hypothesis was found well satisfied even close to localization (\cite{Berthier_E}). 
\begin{equation}\label{eq:Fhom_Fhet}
\dot{\delta d}(x,t) \propto \mathcal{F}(d(x,t),\Delta_0) =  \mathcal{F}_{hom}(d_0,\Delta_0) + \delta\mathcal{F}(\delta d(x,t),d_0,\Delta_0)
\end{equation}
\noindent that provides the damage growth rate as a function of the distribution $d(x,t)$ of damage within the specimen at the prescribed displacement $\Delta_0$. Since our damage model does not allow healing of the material, i.e. $\dot{\delta d}\geq 0$, a more rigorous formulation of the damage  evolution law is $\delta \dot{d}(x,\Delta) \propto \max[0,\mathcal{F}]$. But for the sake of simplicity, this positiveness condition on $\delta \dot{d}$ is not explicitly written in the following. The above equation can finally be written as 
\begin{equation}\label{eq:Fhom_psi}
\dot{\delta d}(x,t) \propto \mathcal{F}_{hom}(d_0,\Delta_0) + \psi(x,d_0,\Delta_0) \ast \delta d(x) - y_\mathrm{c}(x,d(x,t))
\end{equation}
\noindent where the  disorder term $y_\mathrm{c}$ introduced in Eq.~\eqref{eq:Yc} describes the spatial variations of damage energy that depends both on the position in the specimen and the current damage level. The inhomogeneous contribution to the driving force consists of a convolution product accounting for the non-local interactions describing the redistributions of driving force taking place within the specimen after each individual damage event. As shown by~\cite{Berthier_E}, the so-called {\it redistribution kernel} $\psi$ is obtained from a linearization of the driving force around the reference damage level $d_0$ and reads as
\begin{equation}\label{eq:psi}
\psi(x,d_0,\Delta_0)= Y'(d_0,\Delta_0)\alpha_2(x) - Y_c'(d_0)
\end{equation}
\noindent where $\alpha_2(x) =  \exp(-|x|/(2\ell_0))(2\ell_0+|x|)$ denotes the convolution of the interaction function $\alpha(x)$ with itself. In the former expression, the prime represents the derivative with respect to the reference damage level $d_0$. We can verify that an increase $\delta d(x) = \delta d_0 \, \delta(x - x_0)$ of the damage field localized in $x = x_0$ where $\delta(u)$ is the Dirac function results in a variation $\delta \mathcal{F}(x) = \delta d_0 \, \psi(x-x_0)$ of the driving force. In other words, the spatial structure of the load redistribution following an elementary damage event is controlled by the kernel $\psi$, and therefore, by the interaction function $\alpha(x)$ introduced in the definition of the non-local damage variable $\bar{d} = \alpha \ast d$ in Eq.~\eqref{eq:alpha}.

In line with the model of Section~\ref{sec:Sec1} which comprises three contributions to the total damage driving force, namely a slowly increasing load amplitude corresponding to a local contribution, a non-local contribution $\overline {Y}$ emerging from the interactions and finally a disordered term $Y_\mathrm{c}(x,d)$ source of strong non-linearities, we note that the evolution equation~\eqref{eq:Fhom_psi} of the damage field consists of three terms too: (i) A local contribution $\mathcal{F}_{hom}$, which predicts how the average damage level increases with the imposed loading, (ii) a non-local contribution $\delta \mathcal{F} = \psi \ast \delta d$ that couples the damage growth rate in $x$ with the damage level in the other regions of the specimen, and (iii) a quenched noise $y_\mathrm{c}$ that describes the resistance to damage that depends both on the position $x$ in the specimen  and the current damage level $d(x)$. Owing to the dependence of $y_\mathrm{c}$ with the damage level, this equation is strongly non-linear and cannot be solved exactly (see for example~\cite{LeDoussal} for advanced methods of resolution based on functional renormalization group). It turns out that such equations are reminiscent of a broader class of problems referred to as driven disordered elastic interfaces that deal with elastic manifolds driven in disordered media~\citep{Narayan, Barabasi, Leschhorn2, Wiese2}. Following an analogy already used by~\cite{Vandembroucq4}, \cite{Lin} and~\cite{Weiss2}, the damage field behaves then as a uni-dimensional interface of position $\delta d(x)$ driven in a two-dimensional plane $(x,d)$ of disordered resistance $y_\mathrm{c}(x,d)$. The evolution equation~\eqref{eq:Fhom_psi} predicts the interface speed $\dot{\delta d}(x)$ as a function of the current interface position $\delta d(x)$. Note that in this framework, the redistribution kernel $\psi$ describes the elasticity of the interface. Indeed, as one point of the interface located in $x = x_0$ moves forward of an increment $\delta d_0$, the other regions of the interface feels an increase $\delta \mathcal{F}(x) = \delta d_0 \, \psi(x-x_0) > 0$ of the applied driving force.  

\subsection{Scaling behavior of the precursors at fixed distance to failure}~\label{sec:Sec3b}
The behavior of driven disordered elastic interfaces has been thoroughly investigated both theoretically and numerically (see~\cite{Wiese2} for a recent review). Such systems exhibit an intermittent dynamics characterized by micro-instabilities localized both in space and time that can be elegantly described by scaling laws. In particular, the driven elastic interface (and so the damage field that it represents) grows through avalanches, the size, duration and spatial extent of which scale with each other as
\begin{equation}
\left \{
\begin{array}{l c l}
S \sim \ell_\mathrm{x}^{d_\mathrm{f}}  \\
T \sim \ell_\mathrm{x}^{z}.
\end{array}
\right .
\label{Eq_Predictions}
\end{equation}
\noindent The exponents $d_\mathrm{f}$ and $z$ take values that depend only on two features of the interface: Its dimensionality and its elasticity. Here, the interface is a uni-dimensional line while its elasticity is referred to as {\it short-range}, as the redistribution kernel $\psi \sim \alpha_2$ decays exponentially fast with the distance $x$ resulting in  a short-range coupling between the constituting elements of the specimen. Considering 1D interfaces with short-range elasticity, \cite{Rosso9} and \cite{Duemmer2} predicted a fractal dimension $ d_\mathrm{f} \simeq 9/4$ and a dynamic exponent $z \simeq 3/2$. Our numerical investigation of the statistics of precursors carried in Section~\ref{sec:Sec2} led to $ d_\mathrm{f}= 2.35\pm 0.15$ and $z = 1.40 \pm 0.15$, two values that are in good agreement with these theoretically predicted exponents. Moreover, because the scalings are determined at a fixed distance to failure, which can be defined in terms of applied displacement or force, we do not expect variations in the scaling relations between size, duration and spatial extent if considering force as the controlled loading parameter. 
This successful comparison suggests that the theoretical framework of driven disordered elastic interface captures adequately the intermittent dynamics of damage growth, at least at some fixed distance to failure. This confirms the claim made by~\cite{Petri, Zapperi3} that fracture precursors are reminiscent of the critical dynamics of damage growth, even though, as we will show later, the catastrophic failure of the specimen does not correspond to a critical point.  As an additional remark, we can notice that the measured exponent $\beta \simeq 1.5$ characterizing the distribution of avalanche sizes in \eqref{eq:PS_S} corresponds to the \textit{mean-field} exponent, and not the value $\beta \simeq 1.1$ predicted for disordered elastic interfaces with short-range interactions~\citep{rosso7,LePriol}. This observation, consistent with the exponent value reported by~\citet{Raischel} that considered a local fiber model with progressively damaging elements, may be connected to the \textit{non-stationary} nature of depinning model underlying compressive failure, a point thoroughly discussed in the next section.


Before investigating the non-stationary dynamics of damage accumulation in more detail, we would like to emphasize that the nature of interactions between the constituting elements of the specimen, described by the kernel $\psi$ and the function $\alpha$ from which it derives, control both the onset of specimen failure, as thoroughly discussed by~\cite{Berthier_E}, {\it and} the statistical properties of the damage cascades that precede it, as shown here. As a result, the precursory damage activity carry information on the final failure event, an idea that we will use thereafter to design a method of failure prediction from the statistical analysis of damage cascades.

\subsection{Non-stationary evolution equation of the damage field under monotonically increasing loading amplitude}~\label{sec:Sec3c}
We now go a step further and focus on a more realistic scenario of damage growth  as the loading amplitude increases. We seek to explain the non-stationary features of damage evolution, observed numerically in Section 2. As in Section 3.1, we start by decomposing the fields in terms of homogeneous and perturbation contributions. We then explicitly formulate the time dependency to evaluate the velocity, and other characteristic features, of the interface as it is externally driven towards failure. 

We first decompose the damage field in two contributions
\begin{equation}
d(x,t_0) = d_0(t_0) + \delta d(x,t_0) 
\end{equation}
where $d_0(t_0) = \langle d(x,t_0) \rangle_x$ is the mean damage level at the reference time $t_0 = 0$. In the following, as previously, we assume a damage field perturbation that is small in comparison to the average damage level, at all distance to localization. We assume that the damage field perturbations $\delta d \ll d_0$ are small in comparison to the reference damage level $d_0$. This amounts to consider rather small values $\sigma \ll 1$ of disorder amplitude, as damage field fluctuations scales linearly with $\sigma$, as well as rather short amount of time $(t-t_0)$ before the reference time $t_0$ must be updated again.

We can first solve the homogeneous problem that consists in predicting how $d_0$ varies with the imposed displacement $\Delta_0$. The governing equation writes as
\begin{equation}
\mathcal{F}_{hom}(d_0,\Delta_0)=0 \quad \Rightarrow \quad d_0(\Delta_0)
\label{Eq:hom}
\end{equation}
from which we derives the relation between $d_0$ and $\Delta_0$. As a result, we can indifferently use one or the other variable to describe the reference state of the specimen, and thus the distance to catastrophic failure.
 
We are now interested in predicting the evolution of the damage field perturbations. To do so, we assume that the imposed displacement increases linearly with time
\begin{equation}
\Delta(t) = \Delta_0 + \delta \Delta(t) = \Delta_0 + v_{\mathrm{ext}} \, t 
\end{equation}
where the displacement increment $\delta \Delta(t)$ is assumed to be small with respect to the reference displacement $\Delta_0$. This amounts to consider the evolution of the damage field on a rather short amount of time $t - t_0 = t \ll \Delta_0/v_\mathrm{ext}$, before the reference displacement $\Delta_0$ must be updated again.

The equation governing the evolution of $\delta d(x,t)$ derives from a linearization of the evolution law
\begin{equation}
\dot{d}(x,t) \propto \mathcal{F}(d (x,t), \Delta(t)) \quad \Rightarrow \quad \delta \dot{d}(x,t) \propto \mathcal{F}(d_0 + \delta d (x,t), \Delta_0 + \delta \Delta(t))
\end{equation}
where the total damage driving force decomposes into two contributions $\mathcal{F}_\mathrm{hom}(d_0,\Delta_0) + \delta \mathcal{F}(\Delta_0,\delta d(x,t),\delta \Delta(t))$. At the first order in the damage field perturbation $\delta d(x,t)$, the evolution equation of the damage field perturbation writes as
\begin{equation}
\dot{\delta d}(x,t) \propto \frac{\partial \mathcal{F}_\mathrm{hom}}{\partial \Delta}(d_0,\Delta_0) \, \delta \Delta(t) + \psi(d_0,\Delta_0) \ast \delta d(x,t) - y_\mathrm{c}(x,d(x,t)).
\end{equation}
\noindent The last two terms represent the perturbation in the damage driving force under fixed loading amplitude $\Delta(t) = \Delta_0$, which directly derives from Eq.~\eqref{eq:Fhom_psi}. Using the expression of the homogeneous damage driving force $\mathcal{F}_\mathrm{hom}(d_0\Delta_0) = Y(d_0,\Delta_0) - Y_\mathrm{c}(d_0)$ and the expression~\eqref{eq:psi} of the interaction kernel $\psi(d_0,\Delta_0)$, one obtains
\begin{equation}
\dot{\delta d}(x,t) \propto \frac{\partial \mathcal{F}_\mathrm{hom}}{\partial \Delta}(d_0,\Delta_0) \, v_{\mathrm{ext}}t -Y_c'(d_0)\delta d(x) +Y'(d_0,\Delta_0)(\alpha_2\ast \delta d)(x)- y_\mathrm{c}(x,d).
\end{equation}

\noindent Introducing the parameters $\mathcal{K}$, $v_\mathrm{m}$ and $\mathcal{H}$, the evolution equation takes its final form
\begin{equation}\label{eq:EvolEqtV2}
\dot{\delta d}(x,t) \propto \mathcal{K}(\delta) \left[v_\mathrm{m}(\delta) t -\delta d(x,t)\right] + \mathcal{H}(\delta)\displaystyle\int_{\Sigma}\alpha_2(\vert x-x' \vert)\left[\delta d(x',t)-\delta d(x,t)\right]dx' - y_\mathrm{c}(x,d(x,t))
\end{equation}
\begin{equation}\label{eq:deltad}
\mathrm{where} \quad \left \{
  \begin{array}{ll}
\displaystyle \mathcal{K}(\delta) = \mathcal{K}(d_0,\Delta_0) = Y_\mathrm{c}'(d_0)-Y'(d_0,\Delta_0) = - \frac{\partial \mathcal{F}_\mathrm{hom}}{\partial d}(d_0,\Delta_0) \\ \\ 
\displaystyle v_\mathrm{m}(\delta) = v_\mathrm{m}(d_0,\Delta_0) = \frac{\partial  \mathcal{F}_\mathrm{hom}}{\partial \Delta} (d_0,\Delta_0) \, \frac{v_\mathrm{ext}}{\mathcal{K}(d_0,\Delta_0)} \\ \\
\mathcal{H}(\delta) = \mathcal{H}(d_0,\Delta_0) = -Y'(d_0,\Delta_0).
  \end{array}
 \right .
\end{equation}
Here, the dependence of the newly introduced parameters $\mathcal{K}$, $v_\mathrm{m}$ and $\mathcal{H}$ with the sole distance to failure $\displaystyle \delta = \frac{\Delta_\mathrm{c} - \Delta_0}{\Delta_\mathrm{c} - \Delta_\mathrm{el}}$ derives from the homogeneous solution~\eqref{Eq:hom} that is used to replace the reference damage level by its expression $d_0(\Delta_0)$ as a function of $\Delta_0$, and then $\Delta_0$ by $\delta$ in the previous expressions. 
The integral term describes the elastic interactions between different regions of the specimen as the interface moves forward, ultimately controlling the scaling properties of the avalanches, like the relations between their size, their duration and their spatial extent. Separately from this mechanism, the mean temporal variation of these three quantities is controlled by the first term of the equation that describes the external loading conditions. The details of the interaction function thus affect the spatial structure of the avalanches but not the divergence of their size on approaching failure that is robust to the particular choice of interaction function.

\begin{figure}[ht!]
\center
\includegraphics[width = .45\textwidth]{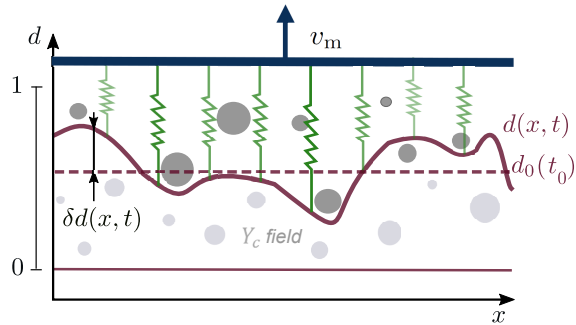} 
\caption{Representation of the damage field evolution by the problem of an elastic interface driven in a random medium. A rigid bar pulls through Hookean springs on an elastic interface of position $d(x,t)$ which describes the damage accumulated in the specimen. The springs stiffness varies with local interface position \-- advanced (resp. nascent) positions experience small (resp. large) stiffness. The interface elasticity, that represents the elastic interactions within the specimen, tries to maintain a homogeneous damage level, while the disorder traps the elastic line in some rough configurations. The competition between elasticity and disorder results in the crackling dynamics with broadly distributed precursors described in Fig.~\ref{fig:Fig1}. The non-stationary behavior of the precursory activity evidenced in Fig~\ref{fig:Fig3} can then be understood as a result of variations in the driving conditions imposed to the interface, both in terms of driving speed and spring stiffness.}
\label{fig:Fig5}
\end{figure}

Under this form, the evolution equation also sheds light on the analogy between the damage field evolution and the problem of an elastic interface driven in a disordered medium, analogy that is schematically illustrated in Fig.~\ref{fig:Fig5}. First of all, the {\it elasticity} of the interface is described by the integral term of Eq.~\eqref{eq:EvolEqtV2}. It provides the driving force distribution along the interface as a function of its geometry $\delta d(x)$. For a flat interface, this contribution is zero and the driving force is constant along the interface. On the contrary, if a point located in $x$ is in advance (resp. behind) with respect to the rest of the interface ($\delta d(x) > \delta d(x')$ for all $x'$), the driving force in $x$ is then smaller (resp. larger) than in the other regions of the interface.\footnote{As discussed in Section~\ref{sec:Sec3a}, this term also provides the spatial structure of the load redistribution taking place after each individual damage event. The spatial structure is described by the kernel $\mathcal{H}(\delta) \alpha_2(x)$ that is equal to the redistribution kernel $\psi$ of Eq.~\eqref{eq:psi} up to a constant $ - Y_\mathrm{c}'(d_0)$. This {\it local} contribution to the damage driving force is accounted for in the first term $\mathcal{K}(\delta) \left[v_\mathrm{m}(\delta) t -\delta d(x,t)\right]$ of the evolution equation~\eqref{eq:EvolEqtV2}, together with the other local contributions.} Second of all, the first term in the evolution equation~\eqref{eq:EvolEqtV2} describes the {\it driving} conditions imposed to the interface. The interface (and thus the damage field it represents) is driven by a rigid bar moving at the speed $v_\mathrm{m}$, as schematically represented in Fig.~\ref{fig:Fig5}. The load transfer from the bar to the interface is ensured by Hookean springs of stiffness $\mathcal{K}$. This particular driving can also be pictured by considering that the interface is trapped by a quadratic potential of width $1/\mathcal{K}$ that moves with a speed $v_\mathrm{m}$. 

Contrary to the analysis carried in Section~\ref{sec:Sec3a}, the parameters involved in Eqs.~\eqref{eq:EvolEqtV2}, like the spring stiffness $\mathcal{K}$ and the speed $v_\mathrm{m}$ at which the interface is driven, now vary with distance to failure, see Eq.~\eqref{eq:deltad}. We then examine these variations and their effect on the precursors statistics. Catastrophic failure taking place at the critical damage level $ d_0 = d_\mathrm{c}$ results from the unstable growth of damage under fixed loading conditions. The failure condition thus writes as $\displaystyle \frac{\partial \mathcal{F}_\mathrm{hom}}{\partial d}(d_\mathrm{c},\Delta_\mathrm{c}) = 0$ that translates, from Eq.~\eqref{eq:deltad}, into a vanishing spring stiffness $\mathcal{K} = 0$ at the failure point $\delta = 0$. This also implies that on approaching failure, the driving speed $v_\mathrm{m} \sim 1/\mathcal{K}$ blows up.

What is the effect of the variations of $v_\mathrm{m}$  and $\mathcal{K}$  on the statistics of precursors ? In standard models of elastic interfaces, the driving controls the distance to the so-called {\it depinning transition}. In particular, considering vanishingly small driving speed and spring stiffness drives the interface in a critical state where the characteristic time $T^\star$ and length scale $\xi$ characterizing the intermittent dynamics diverge. Such a behavior is strikingly similar to the one observed in our simulations and in experiments like the ones of~\cite{Garcimartin, Baro2, Vu}, leading~\cite{Weiss2} and~\cite{Vu} to interpret compressive failure as a critical depinning transition. However, the essential condition of vanishingly small driving speed $v_\mathrm{m} \rightarrow 0$ is not met, as instead, $v_\mathrm{m}$ diverges close to failure (interpreted as a critical point).  As a result, such scenario cannot account neither for the increase of the precursor size observed in our simulations, nor the one reported in experiments, as we will argue later. The role of varying $\mathcal{K}$, which arises from the non-stationary nature of damage and was not considered in~\cite{Weiss2}, is in contrast to standard depinning models. In the following, we present an alternative explanation based on the interpretation of compressive failure as a standard bifurcation.

\subsection{Acceleration of the precursory activity as signature of the unstable nature of compressive failure} 
\label{Sec:Acc}
\begin{figure} 
\center
\includegraphics[width = .9\textwidth]{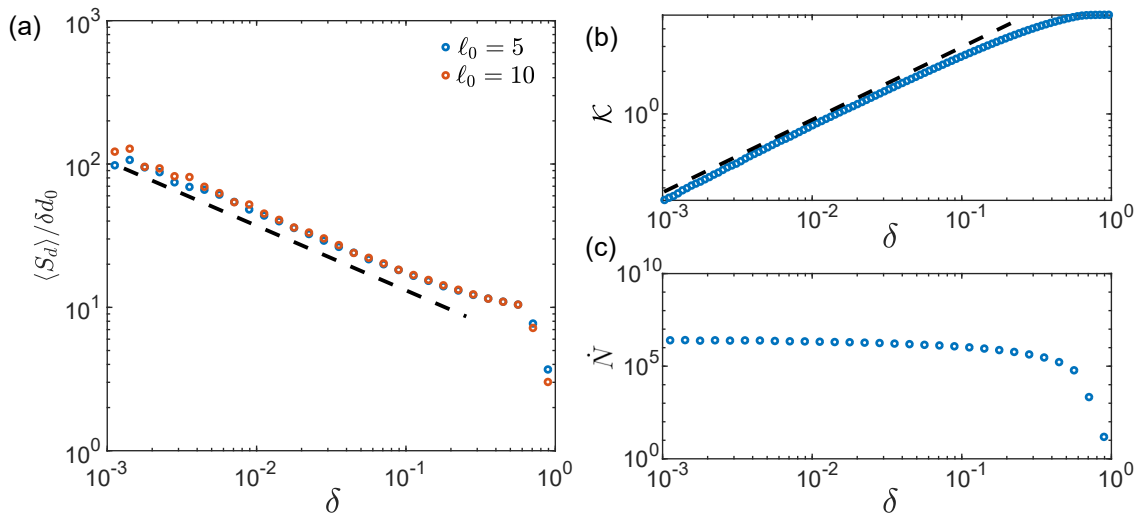} 
\caption{(a) Variations of the precursor size measured in the simulations with the distance to failure. The behavior $\langle S_\mathrm{d} \rangle \propto \langle S \rangle \propto \delta^{-\alpha_\mathrm{s}}$ with an exponent $\alpha_\mathrm{s} = 0.48 \pm 0.05$ (dashed line) is compatible with the square-root singularity predicted in Eq.~\eqref{eq:Pred} from the governing equation~\eqref{eq:EvolEqtV2}; (b) The spring stiffness $\mathcal{K}$ involved in the governing equation~\eqref{eq:EvolEqtV2} and computed from Eq.~\eqref{eq:deltad} is represented as a function of the distance to failure $\delta$. The dashed line $\mathcal{K} \propto \sqrt{\delta}$ confirms that it vanishes as the specimen approaches failure. (c) The rate $\dot{N}$ of precursors is shown as a function of the distance to failure. This behavior is compatible with a standard property of driven elastic interfaces, namely that the rate of avalanches is constant and independent of the interface speed.}
\label{fig:Fig6}
\end{figure}

We first come back to the evolution equation~\eqref{eq:EvolEqtV2} of the damage field and derive from it the evolution of the precursor size with distance to failure. On the one hand, we invoke an important property of driven disordered interfaces, namely that the {\it rate} $\dot{N}$ of precursors is constant and independent of the interface speed~\cite{Wiese2}, a property that is verified in our simulations, as shown in Fig.~\ref{fig:Fig6}(c). This implies that the interface speed is set by the average avalanche size, i.e. $\langle \delta \dot{d} \rangle = \dot{N} \, \langle S \rangle \propto \langle S \rangle$. On the other hand, following the evolution equation~\eqref{eq:EvolEqtV2}, the interface velocity is imposed by the speed $v_\mathrm{m}$ at which the rigid bar moves, so that $ \langle S \rangle \propto v_\mathrm{m}$.\footnote{This property can be deduced by noticing that the first term in the evolution equation~\eqref{eq:EvolEqtV2} of the interface must remain finite, so that $\delta d \propto v_\mathrm{m} \, t$} The expression~\eqref{eq:deltad} of the speed $v_\mathrm{m}$ then leads to
\begin{equation}
\langle S \rangle \sim \dfrac{1}{\mathcal{K}}.
\end{equation}
We now derive the variations of the stiffness $\mathcal{K}$ with the distance to failure $\delta$. As noticed earlier, $\displaystyle \mathcal{K} = - \frac{\partial \mathcal{F}_\mathrm{hom}}{\partial d}(d_0,\Delta_0)$ provides the stability of the process of damage growth for a homogeneous specimen. As a result, $ \mathcal{K}$ vanishes when $\delta = 0$. It turns out that close to failure, $\mathcal{K} \propto \sqrt{\delta}$, a property that is verified in Fig.~\ref{fig:Fig6}(b) and demonstrated in Appendix~A. Therefore, the average avalanche size diverges as
\begin{equation}\label{eq:Pred}
\langle S \rangle \sim \dfrac{1}{\sqrt{\delta}}.
\end{equation}
As shown in Fig.~\ref{fig:Fig6}(a), this behavior accounts for the variations of $\langle S \rangle$ measured in our simulations that shows divergence with an exponent $\alpha_\mathrm{s} \simeq 0.48 \pm 0.05$. Note that while we have considered here displacement-imposed conditions, we can show, following the same logic detailed in Eqs.\eqref{Eq:hom}-\eqref{eq:EvolEqtV2}, that a similar scaling behavior holds under force-imposed conditions, after defining the distance to failure from the distance $F_\mathrm{c} - F$ to force at failure instead of $\Delta_\mathrm{c} - \Delta$ as in Eq.~\eqref{eq:delta} 

We are now in position to explain the divergence of the characteristic time and length scale of the precursors observed close to failure in our simulations. We focus first on the size of the largest precursors $S^\star$ that can be related to $\langle S \rangle$ from the expression~\eqref{eq:PS_S} of the precursor size distribution,
\begin{equation}
\langle S \rangle \propto (S^\star)^{2-\beta} \propto \sqrt{S^\star} 
\end{equation}
where the exponent $\beta \simeq 1.5$ is inferred from the simulations. Combined with the square-root divergence~\eqref{eq:Pred} of the average precursor size, one obtains
\begin{equation}
S^\star \sim 1/\delta
\end{equation}
that accounts for the behavior reported in Eq.~\eqref{eq:S_vs_delta} for the largest precursors. Using finally the scaling relations $T^\star \sim (S^\star)^{z/d_\mathrm{f}}$ and $\ell_\mathrm{x}^\star \sim (S^\star)^{1/d_\mathrm{f}}$ between the duration, the spatial extent and the size of the largest precursors, we predict 
\begin{equation}
\left \{
\begin{array}{lcl} \vspace{5pt}
T^\star & \sim & 1/\delta^{\phi} \quad \mathrm{with} \quad \phi = z/d_\mathrm{f} \simeq 2/3 \\
\ell_\mathrm{x}^\star & \sim & 1/\delta^{\kappa} \quad \mathrm{with} \quad \kappa = 1/d_\mathrm{f} \simeq 4/9,
\end{array}
\right .
\end{equation}
two equations that account for the numerical observations of Eqs.~\eqref{eq:ellx_vs_delta} and~\eqref{eq:T_vs_delta}.

The ability of our model to describe quantitatively the non-stationary dynamics of failure precursors observed in our simulations calls for a few comments. First, the evolution equation~\eqref{eq:EvolEqtV2} derived from the model described in Section~\ref{sec:Sec1a} explicitly connects the non-stationnary dynamics of precursors and the loss of stability of the damage growth process, a feature that was not evident from the simulation results. Consequently, the divergence of the duration, size and spatial extent of the damage cascades close to failure does not derive from the depinning transition of the elastic manifold describing the damage field.\footnote{Note that in such a scenario, the exponents $\phi$, $\kappa$ and $\gamma$ characterizing the divergences $T^\star  \sim  1/\delta^{\phi}$, $\ell_\mathrm{x}^\star  \sim  1/\delta^{\kappa}$ and $S^\star  \sim  1/\delta^{\gamma}$ are set by the so-called {\it correlation length} exponent $\nu \simeq 4/3$ leading to $\phi = \nu \simeq 4/3$, $\kappa = \nu \, z \simeq 2$ and $\gamma = \nu \, d_\mathrm{f} \simeq 3$ incompatible with the exponents measured in the simulations.}. Instead the elastic interface that describes the damage field is {\it driven away} from the critical point. This phenomenon takes place through the divergence of the driving speed and thus the damage rate, a behavior that is reminiscent of the presence of a bifurcation point in $\delta = 0$ leading to the full failure of the specimen, irrespective of the presence of material disorder. This conclusion is in line with the ones drawn by~\cite{Zapperi5} who investigated failure in disordered materials by micro-cracking using a discrete approach referred to as random fuse model. Such a description may also lead to a divergence of the precursor size close to failure, without the system reaching a critical state. Instead, the failure process can be described as a standard bifurcation for which the disorder is irrelevant. Such mechanism, referred to as {\it sweeping of an instability}, encompasses many physical systems including ferromagnetic and granular materials~\citep{Sornette}.

In fracture, the absence of critical behavior at the failure point has practical consequences. Indeed, the exponents characterizing the precursor evolution close to failure do not derive from the theory of critical phenomena, but instead, can be predicted from standard bifurcation theory. As a result, the exponent that describes the divergence of the avalanche size close to failure is expected to keep a constant value $\alpha_\mathrm{s} = 1/2$, irrespective of the range of the interactions as well as the system dimension. This means that in principle we expect this exponent to hold  in models of more realistic systems, including 3D elasto-damageable materials for which elasticity leads to long-range interactions. Yet, we note that the divergence of the precursors size characterized by $\alpha_\mathrm{S} = 0.5$ observed in our simulations and explained in our model differs from the value $\alpha_\mathrm{S} = 1.0$ measured in simulations considering Eshelby-like interaction functions~\citep{Girard} and in molecular dynamics simulations~\citep{Karimi}. We now harness this property to design a methodology that predicts the residual lifetime of materials and structures from the statistical analysis of the failure precursors.

 
\section{Predicting failure from the statistics of precursors: A numerical proof of concept}
As an application of our theoretical findings, we now would like to discuss how the acceleration of precursory activity as a signature of an impending instability can be used in structural health monitoring. Damage accumulation results in the progressive loss of stiffness of structures, ultimately threatening their mechanical integrity. Tracking damage and its evolution inside structures is a very challenging task.\footnote{X-ray tomography, although used to characterize damage inside small specimens, see {\it e.g.}~\cite{kandula2019} and~\cite{cartwright2020}, is not adapted to study large structures.} As a result, acoustic emissions, that accompany damage growth and can be recorded by piezoelectric sensors, have been used as a ready-made source of information for monitoring their mechanical health. Laboratory experiments show that the amplitude and frequency of the acoustic bursts increase on approaching failure~\citep{Garcimartin, Baro, Vu}. Taking inspiration from seismology, such an increase has been described using the so-called time-reversed Omori-law\footnote{The standard Omori-law describes the rate of the after-shocks following large earthquakes~\citep{hirata1987a, hirata1987b, Baro, Baro2, salje2017}. It follows Eq.~\eqref{Eq_InvOL} with the noticeable difference that $t-t_\mathrm{c}$ is used instead of $t_\mathrm{c} - t$.}
\begin{equation}
\centering
A(t) = \frac{A_\mathrm{o}} {(t_\mathrm{c} - t)^{p}}.
\label{Eq_InvOL}
\end{equation}
In this expression, $A(t)$ is the rate of acoustic bursts recorded at time $t$ before catastrophic failure that takes place at time $t_\mathrm{c}$. $A_\mathrm{o}$ is a proportionality constant while the exponent $p$ is found to be close to one~\citep{Baro}, even though it has been shown to vary with strain rate and temperature~\citep{ojala2004}. Inspired by procedures used for predicting earthquakes, landslides and volcanic eruptions ~\citep{voight1988, kilburn1998, Sornette, bell2013, bell2018} where similar power-law divergence of precursors have been reported, it has recently been proposed that such a scaling behavior could be used for monitoring the mechanical health of structures~\citep{Mayya}, following the idea of~\cite{Anifrani}.

As a numerical proof of concept,  we now use the precursors signals obtained by numerical implementation of our model described in Section~\ref{sec:numerical} to  perform a retrospective failure prediction. Our model does not provide the acoustic signal accompanying damage growth. Hence, we assume that the size $S$ (in energy) of damage precursors can be inferred from the acoustic signal recorded from the real structure. As a result, we consider the time-series $S(t)$  that follows a scaling behavior similar to Eq.~\eqref{Eq_InvOL} as input data for failure prediction. We consider that damage starts to grow at $t = 0$ (corresponding to $\Delta = \Delta_\mathrm{el}$, {\it i.e.} $\delta =1$) and specimen failure occurs at $t = t_\mathrm{c}$ (corresponding to $\Delta = \Delta_\mathrm{c}$, {\it i.e.} $\delta = 0$). Assuming that the displacement is increased linearly with time, the distance to failure writes then as $\delta = (t_\mathrm{c} - t)/t_\mathrm{c}$. The scaling law~\eqref{eq:Pred} then rewrites as
\begin{equation}
\centering
\langle S \rangle = \frac{S_\mathrm{o}}{(t_\mathrm{c} - t)^{\alpha_s}},
\label{eq.predict}
\end{equation}
\noindent where $\alpha_\mathrm{s} = 1/2$. This equation is developed in the form
\begin{equation}
\centering
\langle S \rangle^{1/\alpha_\mathrm{s}} \, t   =  t_\mathrm{c} \ \langle S \rangle^{1/\alpha_\mathrm{s}}  - S_\mathrm{o} \label{eq.predict2}
\end{equation}
that follows $Y = t_\mathrm{s} \, X - S_\mathrm{o}$ where both $Y = \langle S \rangle^{1/\alpha_\mathrm{s}} \, t$ and $X = \langle S \rangle^{1/\alpha_\mathrm{s}}$ can be computed at different time steps during the monitoring so that the failure time $t_\mathrm{c}$ can be subsequently predicted from a linear fit.

\begin{figure}[ht!]
\center
\includegraphics[width = 0.75\textwidth]{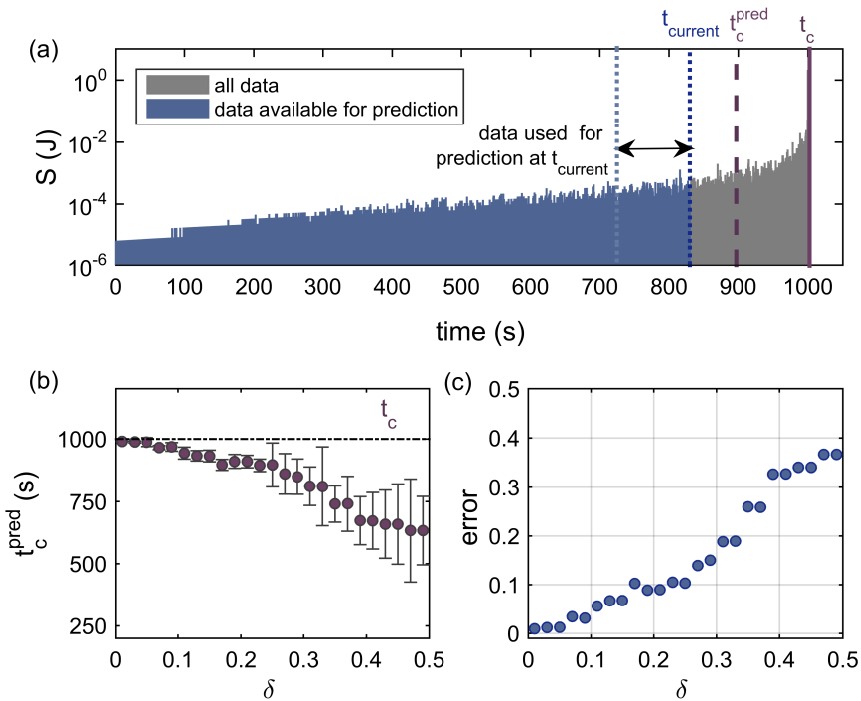} 
\caption{(a) Evolution of avalanche energy, $S$ as a function of the time. An example case of the prediction $t_\mathrm{c}^{\mathrm{pred}}$ for a value of $t_{\mathrm{current}}$ is also presented.(b) Predicted values of time to failure and (c) the error in prediction ($(t_\mathrm{c} - t_\mathrm{c}^{\mathrm{pred}})/t_\mathrm{c}$) at different instances of $\delta = (t_\mathrm{c} - t)/t$.}
\label{fig:Fig7}
\end{figure}

In practice, the implementation of the method is as follows. We assume that the process of damage spreading takes $t_\mathrm{c} = 1000~\mathrm{s}$ until the specimen fails, as shown in Fig.~\ref{fig:Fig7}(a). This amounts to assume a loading rate $\dot{\Delta} = (\Delta_\mathrm{c} - \Delta_\mathrm{el})/1000$. The mean precursors size $\langle S \rangle$ is obtained from an average over non-overlapping time windows of $20~\mathrm{s}$, a duration sufficiently small to accurately capture the acceleration of precursory activity. The monitoring of the damage activity is performed at time $t_\mathrm{current}$. We then record damage events in the immediate history of $t_\mathrm{current}$, {\it i.e.} during $100~\mathrm{s}$ before $t_\mathrm{current}$. Equation~\eqref{eq.predict2} is then used to fit the data and predicts the failure time. The prediction $t_\mathrm{c}^\mathrm{pred}$ is shown in Fig.~\ref{fig:Fig7}(b) as a function of the residual lifetime $\delta = (t_\mathrm{c} - t_\mathrm{current})/t_\mathrm{c}$ when the monitoring is performed. It is also compared with the actual failure time $t_\mathrm{c} = 1000~\mathrm{s}$. The relative error $(t_\mathrm{c}-t_\mathrm{c}^\mathrm{pred})/t_\mathrm{c}$ on the prediction is shown in Fig.~\ref{fig:Fig7}(c). It rapidly decreases as the prediction is made closer to final failure. The predictions are within 10\% of the actual failure time when the remaining lifetime is less than 25\%.

Interestingly, the proposed methodology is conservative, predicting systematically failure time {\it smaller} than $t_\mathrm{c}$. Note also that it does not require monitoring from the beginning of the damage accumulation phase, unlike cumulative record based predictions~\citep{Carpinteri3, Abdelrahman}. Accurate residual life-assessment can be inferred at any time from data recorded on a rather short amount of time whereas empirical law based approaches require detailed experimental protocols for measuring various material parameters in the lab before providing reliable predictions for the structure~\citep{Godin}. In practice, working with the raw acoustic signal emitted by the structure and not the mechanical precursors $S(t)$ renders the post-processing more complex. However, several strategies have been proposed by~\cite{Mayya} to circumvent these difficulties and they are now tested in laboratory experiments where both the acoustic precursors and the mechanical precursors can be recorded.


\section{Conclusions}
Compressive failure of quasi-brittle solids results from the complex evolution of a large number of dissipative events, such as micro-cracking, in interaction with each other. Here, we examine such a collective dynamics and its connection with final failure through the statistical analysis of the damage precursors taking place during the phase of damage accumulation. In practice, we analyze the intermittent mechanical response of a disordered elasto-damageable specimen under compression as predicted from a minimal thermodynamically-consistent damage model developed by~\cite{Berthier_E}. This model captures the co-action of material disorder and elastic interactions conveyed by the non-local stress distribution taking place in the specimen after each individual damage event. The major results of our study are the following:
\begin{enumerate}
\item The proposed description predicts not only localization and catastrophic failure, as shown by~\cite{Berthier_E}, but also captures also qualitatively the main features of the jerky evolution of damage observed experimentally in quasi-brittle solids under compression~\citep{Petri, Garcimartin, Guarino2, Rosti, Baro, Vu}. Damage grows through bursts the size, duration and spatial extent of which are related to each other by scaling laws. As the specimen is driven towards failure, these quantities increase following power law relationships with the distance to final failure.
\item To rationalize these numerical observations, we then investigate theoretically the intermittent growth of damage. We derive an evolution equation of the damage field within the specimen that captures quantitatively all the statistical properties of the precursors evidenced in our simulations. It also elicits the connection between compressive failure and the theoretical framework of elastic interfaces driven in disordered media used in former studies~\citep{Weiss2} to represent the damage field. This mapping accounts for the scaling relations between the size, the duration and the spatial extent of the failure precursors. The exponents involved in these scaling relations are critical exponents emerging from the so-called depinning transition, a second order phase transition taking place when the driving speed of the elastic interface goes to zero. However, and contrary to the discussion by~\cite{Weiss2} and~\cite{Vu}, the increase of these quantities close to failure does not emerge from the depinning transition. Instead, it results from the divergence of the damage rate (that translates into the divergence of the interface driving speed), a feature that arises from the presence of a standard bifurcation at the compressive failure threshold~\cite{Berthier_E,Dansereau}. This new explanation for the non-stationary features of the damage evolution accounts for our numerical observations and in particular for the square-root divergence of the precursor size with distance to failure. 
\item Last but not least, we finally harness the universal statistical properties of failure precursors to design an innovative technique of structural health monitoring inspired from the method proposed by~\cite{Anifrani}. Using our numerical data, we show that short time-series of precursors size available ahead of final failure can be processed to predict the residual lifetime with accuracy. As the proposed methodology is based on the physics of damage spreading in quasi-brittle solids, we expect its predictions to be more robust than the methods used currently in structural health monitoring.
\end{enumerate}
Finally, we would like to discuss how the results derived in this study for a model elasto-damageable solid may apply to more realistic situations. The nature of the interactions in real materials being different, {\it i.e.} long-ranged while we here considered short-range interactions, the exponents characterizing the relations between the size, the duration and the spatial extent are expected to differ. Nonetheless, the scaling relations between these quantities are expected to hold and the exponents can be predicted from the theory of driven disordered interfaces using the appropriate interaction kernel~\citep{Dansereau} and specimen dimensionality \--- 1D {\it vs.} 2/3D. The scaling laws characterizing the divergence of these quantities with the distance to failure are shown to be independent of the dimensionality of the specimen and the nature of the elastic interactions. Hence, the evolution of precursors towards failure is expected to be characterized by the simple and robust scaling laws deduced from the stability analysis, reinforcing further the robustness of the proposed methodology of structural health monitoring~\citep{Mayya}. Interestingly, the depinning scenario discussed in \citet{Weiss2} was used to explain the experimental observations of statistical size effects of strength in quasi-brittle materials, in particular the non-vanishing strength at large size. Building on the new understanding of the precursors derived from the present study, we are now revisiting the experimentally observed size effect. In addition, we aim to apply these ideas to explain quantitatively the scaling properties of precursors observed in real materials through experiments on model elasto-damageable
solids.


In summary, the minimal model of damage growth used in this study paves the way for a deep understanding of the physics of compressive failure. Indeed, it could describe, at least qualitatively, all its salient features, namely localization, intermittency and non-stationary dynamics. It turned out that these different phenomena emerged from different mechanisms, albeit connected to each other through the long-range elastic interactions within the specimen that control both the statistics of damage cascades during the accumulation phase and the final failure. The scenario derived from our work suggests that damage precursors are simple by-products of the damage growth, that could potentially be shut down if the material disorder could be shut down too. As a result, they can't explain the final failure, but instead, constitute ready-made signals informing about the ongoing damage accumulation. This idea has major implications in structural health monitoring, in particular when it comes to decipher the acoustic emission data to predict the residual lifetime of structures. The next challenges lie certainly in understanding in depth the complex relationship between the acoustic precursors and the mechanical ones investigated in this study as well as investigating experimentally at the microstructural scale the complex spatio-temporal evolution of damage in 2D and 3D elasto-damageable materials in order to challenge and refine the proposed scenario.

\section*{Acknowledgements}
The authors gratefully acknowledge financial support from Sorbonne Universit\'e through the Emergence grant for the research project From damage spreading to failure in quasi-brittle materials as well as CNRS and Satt-Lutech through the tech transfer project - Development of a technology of predictive maintenance for materials and structures under compression.

\appendix

\section{Evolution of spring stiffness $\mathcal{K}$ close to failure}

In this appendix,  we describe the effect of the non-stationarity of  loading on the stability of damage evolution given by  stiffness $\mathcal{K}(d_0,\Delta_0)$ of the springs connecting the rigid bar to the interface. In the mean-field limit, we have
\begin{equation}
\mathcal{K}(d_0,\Delta_0) = -\frac{\partial{\mathcal{F}_{\mathrm{hom}}}}{\partial{d}}(d_0,\Delta_0) = Y_{c0}\eta + \frac{1}{2}\Delta^2_{0}k''(d_0)
\end{equation}

For instances when $d_0 \rightarrow d_c$ corresponding to $\Delta_0 \rightarrow \Delta_c$, the stiffness $\mathcal{K}$ can be expressed by a series expansion around $(d_c,\Delta_c)$ as follows.
\begin{equation}
\mathcal{K}(d_0,\Delta_0) = -\frac{\partial{\mathcal{F_{\mathrm{hom}}}}}{\partial{d}} (d_c, \Delta_c) - \frac{\partial^2{\mathcal{F}_{\mathrm{hom}}}}{\partial{d^2}} (d_c - d_0) - \frac{\partial^2{\mathcal{F}_{\mathrm{hom}}}}{\partial{d}\partial{\Delta}}(\Delta_c - \Delta_0)
\label{eq:K}
\end{equation}

\noindent where, $d_c$ and $\Delta_c$ are the localization threshold and critical loading amplitude, respectively. From \citet{Berthier_E}, the damage evolution was understood to become unstable at localization threshold implying $\frac{\partial{\mathcal{F_{\mathrm{hom}}}}}{\partial{d}} (d_c, \Delta_c) = 0$ leaving the second and third terms of Eq.~\eqref{eq:K}. Further, quantities $(d_c - d_0)$ and $(\Delta_c - \Delta_0)$ in these terms are measures of the distance to failure in terms of damage levels and loading amplitude, respectively. We now proceed to obtain a relation between these two quantities  to decipher the effect on $\mathcal{K}$.
The damage driving force in the mean-field limit $\mathcal{F_{\mathrm{hom}}}$ for damage levels $d_0\rightarrow d_{c}$ follows as
\begin{equation}
\centering
\begin{split}
\mathcal{F_{\mathrm{hom}}}(d_0,\Delta) = &~\mathcal{F_{\mathrm{hom}}}(d_{c},\Delta_c) + \frac{\partial{\mathcal{F_{\mathrm{hom}}}}}{\partial{d}}(d_c - d_0) + \frac{\partial{\mathcal{F_{\mathrm{hom}}}}}{\partial{\Delta}}(\Delta_c - \Delta_0)  \\
& + \frac{1}{2}\frac{\partial^2{\mathcal{F_{\mathrm{hom}}}}}{\partial{d^2}}(d_c - d_0)^2 +  \frac{1}{2}\frac{\partial^2{\mathcal{F_{\mathrm{hom}}}}}{\partial{\Delta^2}}(\Delta_c - \Delta_0)^2
\end{split}
\end{equation}

From governing equation, we obtain both $\mathcal{F_{\mathrm{hom}}}(d_0,\Delta)$ and $\mathcal{F_{\mathrm{hom}}}(d_{loc},\Delta_c) = 0$. Re-invoking the failure criterion,  $\frac{\partial{\mathcal{F_{\mathrm{hom}}}}}{\partial{d}} (d_c, \Delta_c) = 0$, we obtain,

\begin{equation}
\centering
(d_c - d_0)^2 \approx A_0 (\Delta_c - \Delta_0) \longrightarrow (d_c - d_0) \sim (\Delta_c - \Delta_0)^{0.5}
\label{eq:d_Delta}
\end{equation}  

\noindent where, $A_0$ is a constant and the term containing $(\Delta_c - \Delta_0)^2$  is neglected for being small when $\Delta_0 \rightarrow \Delta_c$. Comparing Eqs.~\eqref{eq:K} and \eqref{eq:d_Delta}, we obtain
\begin{equation}
\mathcal{K}(d_0,\Delta_0) \sim \sqrt{\delta}
\end{equation}

\noindent where, $\delta \sim (\Delta_c - \Delta_0)$ and the third term in Eq.~\eqref{eq:K} is neglected for being small when $\Delta \rightarrow \Delta_c$.

\section{Summary of exponents and scaling laws.}
We here summarize the scaling laws relating exponents and compare measured  values to 
scaling-inferred and theory prediction values. In the table, the avalanche size $S_\mathrm{d}$ can be replaced by $S$, owing to their linear relationship.

\begin{center}
 \begin{tabular}{| c | c | c | c | l |} 
 \hline
Equation & Relation & Exponent & Measure & Scaling  \\ [0.5ex] 
 \hline\hline
 (8) & $\ell_x\sim S_d^{1/d_f}$ & $d_f$ & $2.35\pm0.15$ & (20) Theory: $9/4 = 2.25$ \\ 
 \hline
 (9) & $T\sim S_d^{z/d_f}$ & $z$ & $1.40\pm0.15$ & (20) Theory: $3/2 = 1.5$ \\
\hline
 (12) & $S^\ast_d\sim 1/\delta^\gamma$ & $\gamma$ & $1.00\pm0.10$ & (30) \& (31)$ \Rightarrow$(32) \\
 &  &  &  & $\simeq 1.0$ \\
 \hline
 (13) & $T^\ast \sim 1/\delta^\phi$ &  $\phi$ & $0.52\pm 0.10$ & (9), (12) \& (13)$ \Rightarrow$(14), (9) \& (32)$\Rightarrow$(33)\\
  &  &  &  & $\simeq 0.6$ \\
  \hline
 (14) & $\ell_x^\ast \sim 1/\delta^\kappa$ &  $\kappa$ & $0.37\pm 0.10$ & (8), (12) \& (13)$\Rightarrow$(14), (8) \& (32)$\Rightarrow$(33)\\
  &  &  &  & $\simeq 0.42$ \\
  \hline
 (13) & $\xi\sim 1/\delta^\rho$ & $\rho$ & $0.35\pm 0.10$ &  \\   
\hline
 (30) & $\langle S \rangle \sim 1/\delta^{\alpha_s}$ & $\alpha_s$ & $0.48\pm 0.05$ & Theory: $1/2 = 0.50$ \\
\hline
\end{tabular}
\end{center}

\bibliographystyle{elsarticle-harv}

\begin{thebibliography}{89}
\expandafter\ifx\csname natexlab\endcsname\relax\def\natexlab#1{#1}\fi
\providecommand{\url}[1]{\texttt{#1}}
\providecommand{\href}[2]{#2}
\providecommand{\path}[1]{#1}
\providecommand{\DOIprefix}{doi:}
\providecommand{\ArXivprefix}{arXiv:}
\providecommand{\URLprefix}{URL: }
\providecommand{\Pubmedprefix}{pmid:}
\providecommand{\doi}[1]{\href{http://dx.doi.org/#1}{\path{#1}}}
\providecommand{\Pubmed}[1]{\href{pmid:#1}{\path{#1}}}
\providecommand{\bibinfo}[2]{#2}
\ifx\xfnm\relax \def\xfnm[#1]{\unskip,\space#1}\fi
\bibitem[{Abdelrahman et~al.(2014)Abdelrahman, ElBatanouny and
  Ziehl}]{Abdelrahman}
\bibinfo{author}{Abdelrahman, M.}, \bibinfo{author}{ElBatanouny, M.K.},
  \bibinfo{author}{Ziehl, P.H.}, \bibinfo{year}{2014}.
\newblock \bibinfo{title}{Acoustic emission based damage assessment method for
  prestressed concrete structures: Modified index of damage}.
\newblock \bibinfo{journal}{Engineering Structures} \bibinfo{volume}{60},
  \bibinfo{pages}{258--264}.
\bibitem[{Alava et~al.(2006)Alava, Nukala and Zapperi}]{Alava2}
\bibinfo{author}{Alava, M.J.}, \bibinfo{author}{Nukala, P.K.},
  \bibinfo{author}{Zapperi, S.}, \bibinfo{year}{2006}.
\newblock \bibinfo{title}{Statistical models of fracture}.
\newblock \bibinfo{journal}{Adv. Phys.} \bibinfo{volume}{55},
  \bibinfo{pages}{349--476}.
\bibitem[{Amitrano(2006)}]{Amitrano4}
\bibinfo{author}{Amitrano, D.}, \bibinfo{year}{2006}.
\newblock \bibinfo{title}{Rupture by damage accumulation in rocks}.
\newblock \bibinfo{journal}{Int. J. Frac.} \bibinfo{volume}{139},
  \bibinfo{pages}{369--381}.
\bibitem[{Anifrani et~al.(1995)Anifrani, Floch'H, Sornette and
  Souillard}]{Anifrani}
\bibinfo{author}{Anifrani, J.C.}, \bibinfo{author}{Floch'H, C.M.L.},
  \bibinfo{author}{Sornette, D.}, \bibinfo{author}{Souillard, B.},
  \bibinfo{year}{1995}.
\newblock \bibinfo{title}{Method for the predictive determination of load of a
  structure at rupture}.
\bibitem[{Ashby and Sammis(1990)}]{Ashby}
\bibinfo{author}{Ashby, M.F.}, \bibinfo{author}{Sammis, C.G.},
  \bibinfo{year}{1990}.
\newblock \bibinfo{title}{The damage mechanics of brittle solids in
  compression}.
\newblock \bibinfo{journal}{Pure Appl. Geophys.} \bibinfo{volume}{133},
  \bibinfo{pages}{489--521}.
\bibitem[{Barab\'asi and Stanley(1995)}]{Barabasi}
\bibinfo{author}{Barab\'asi, A.L.}, \bibinfo{author}{Stanley, H.E.},
  \bibinfo{year}{1995}.
\newblock \bibinfo{title}{Fractal concepts in surface growth}.
\newblock \bibinfo{publisher}{Cambridge University Press}.
\bibitem[{Bar{\'o} et~al.(2013)Bar{\'o}, Corral, Illa, Planes, Salje, Schranz,
  Soto-Parra and Vives}]{Baro}
\bibinfo{author}{Bar{\'o}, J.}, \bibinfo{author}{Corral, A.},
  \bibinfo{author}{Illa, X.}, \bibinfo{author}{Planes, A.},
  \bibinfo{author}{Salje, E.K.H.}, \bibinfo{author}{Schranz, W.},
  \bibinfo{author}{Soto-Parra, D.E.}, \bibinfo{author}{Vives, E.},
  \bibinfo{year}{2013}.
\newblock \bibinfo{title}{Statistical similarity between the compression of a
  porous material and earthquakes}.
\newblock \bibinfo{journal}{Phys. Rev. Lett.} \bibinfo{volume}{110},
  \bibinfo{pages}{088702}.
\bibitem[{Bar{\'o} et~al.(2018)Bar{\'o}, Dahmen, Davidsen, Planes, Castillo,
  Nataf, Salje and Vives}]{Baro2}
\bibinfo{author}{Bar{\'o}, J.}, \bibinfo{author}{Dahmen, K.},
  \bibinfo{author}{Davidsen, J.}, \bibinfo{author}{Planes, A.},
  \bibinfo{author}{Castillo, P.}, \bibinfo{author}{Nataf, G.F.},
  \bibinfo{author}{Salje, E.K.H.}, \bibinfo{author}{Vives, E.},
  \bibinfo{year}{2018}.
\newblock \bibinfo{title}{Experimental evidence of accelerated seismic release
  without critical failure in acoustic emissions of compressed nanoporous
  materials}.
\newblock \bibinfo{journal}{Phys. Rev. Lett.} \bibinfo{volume}{120},
  \bibinfo{pages}{145501}.
\bibitem[{Bazant(1994)}]{Bazant9}
\bibinfo{author}{Bazant, Z.P.}, \bibinfo{year}{1994}.
\newblock \bibinfo{title}{Nonlocal damage theory based on micromechanics of
  crack interactions}.
\newblock \bibinfo{journal}{J. Eng. Mech.} \bibinfo{volume}{120},
  \bibinfo{pages}{593--617}.
\bibitem[{Bell(2018)}]{bell2018}
\bibinfo{author}{Bell, A.F.}, \bibinfo{year}{2018}.
\newblock \bibinfo{title}{Predictability of landslide timing from
  quasi-periodic precursory earthquakes}.
\newblock \bibinfo{journal}{Geophysical Research Letters} \bibinfo{volume}{45},
  \bibinfo{pages}{1860--1869}.
\bibitem[{Bell et~al.(2013)Bell, Naylor and Main}]{bell2013}
\bibinfo{author}{Bell, A.F.}, \bibinfo{author}{Naylor, M.},
  \bibinfo{author}{Main, I.G.}, \bibinfo{year}{2013}.
\newblock \bibinfo{title}{The limits of predictability of volcanic eruptions
  from accelerating rates of earthquakes}.
\newblock \bibinfo{journal}{Geophysical Journal International}
  \bibinfo{volume}{194}, \bibinfo{pages}{1541--1553}.
\bibitem[{Berthier(2015)}]{Berthier_E3}
\bibinfo{author}{Berthier, E.}, \bibinfo{year}{2015}.
\newblock \bibinfo{title}{Quasi-brittle failure of heterogeneous materials :
  Damage statistics and localization}.
\newblock Ph.D. thesis. Universit\'e Pierre et Marie Curie.
\bibitem[{Berthier et~al.(2017)Berthier, D\'emery and Ponson}]{Berthier_E}
\bibinfo{author}{Berthier, E.}, \bibinfo{author}{D\'emery, V.},
  \bibinfo{author}{Ponson, L.}, \bibinfo{year}{2017}.
\newblock \bibinfo{title}{Damage spreading in quasi-brittle heterogeneous
  materials: I. localization and failure}.
\newblock \bibinfo{journal}{J. Mech. Phys. Solids} \bibinfo{volume}{102},
  \bibinfo{pages}{101--124}.
\bibitem[{Bigoni(2012)}]{Bigoni}
\bibinfo{author}{Bigoni, D.}, \bibinfo{year}{2012}.
\newblock \bibinfo{title}{Nonlinear solid mechanics: Bifurcation theory and
  material instability}.
\newblock \bibinfo{publisher}{Cambridge University Press},
  \bibinfo{address}{Cambridge, England}.
\bibitem[{Carpinteri et~al.(2007)Carpinteri, Lacidogna and Pugno}]{Carpinteri3}
\bibinfo{author}{Carpinteri, A.}, \bibinfo{author}{Lacidogna, G.},
  \bibinfo{author}{Pugno, N.}, \bibinfo{year}{2007}.
\newblock \bibinfo{title}{Structural damage diagnosis and life-time assessment
  by acoustic emission monitoring}.
\newblock \bibinfo{journal}{Eng. Frac. Mech.} \bibinfo{volume}{74},
  \bibinfo{pages}{273--289}.
\bibitem[{Cartwright-Taylor et~al.(2020)Cartwright-Taylor, Main, Butler,
  Fusseis, Flynn and King}]{cartwright2020}
\bibinfo{author}{Cartwright-Taylor, A.}, \bibinfo{author}{Main, I.G.},
  \bibinfo{author}{Butler, I.B.}, \bibinfo{author}{Fusseis, F.},
  \bibinfo{author}{Flynn, M.}, \bibinfo{author}{King, A.},
  \bibinfo{year}{2020}.
\newblock \bibinfo{title}{Catastrophic failure: How and when? insights from 4-d
  in situ x-ray microtomography}.
\newblock \bibinfo{journal}{Journal of Geophysical Research: Solid Earth}
  \bibinfo{volume}{125}, \bibinfo{pages}{e2020JB019642}.
\bibitem[{Chopin et~al.(2018)Chopin, Bhaskar, Jog and Ponson}]{Chopin5}
\bibinfo{author}{Chopin, J.}, \bibinfo{author}{Bhaskar, A.},
  \bibinfo{author}{Jog, A.}, \bibinfo{author}{Ponson, L.},
  \bibinfo{year}{2018}.
\newblock \bibinfo{title}{Depinning dynamics of crack fronts}.
\newblock \bibinfo{journal}{Phys. Rev. Lett.} \bibinfo{volume}{121},
  \bibinfo{pages}{235501}.
\bibitem[{Dansereau et~al.(2019)Dansereau, D{\'e}mery, Berthier, Weiss and
  Ponson}]{Dansereau}
\bibinfo{author}{Dansereau, V.}, \bibinfo{author}{D{\'e}mery, V.},
  \bibinfo{author}{Berthier, E.}, \bibinfo{author}{Weiss, J.},
  \bibinfo{author}{Ponson, L.}, \bibinfo{year}{2019}.
\newblock \bibinfo{title}{Collective damage growth controls fault orientation
  in quasibrittle compressive failure}.
\newblock \bibinfo{journal}{Phys. rev. lett.} \bibinfo{volume}{122},
  \bibinfo{pages}{085501}.
\bibitem[{Davidsen et~al.(2007)Davidsen, Stanchits and Dresen}]{Davidsen}
\bibinfo{author}{Davidsen, J.}, \bibinfo{author}{Stanchits, S.},
  \bibinfo{author}{Dresen, G.}, \bibinfo{year}{2007}.
\newblock \bibinfo{title}{Scaling and universality in rock fracture}.
\newblock \bibinfo{journal}{Phys. Rev. Lett.} \bibinfo{volume}{98},
  \bibinfo{pages}{125502}.
\bibitem[{Deschanel et~al.(2006)Deschanel, Vanel, Vigier, Godin and
  Ciliberto}]{Deschanel}
\bibinfo{author}{Deschanel, S.}, \bibinfo{author}{Vanel, L.},
  \bibinfo{author}{Vigier, G.}, \bibinfo{author}{Godin, N.},
  \bibinfo{author}{Ciliberto, S.}, \bibinfo{year}{2006}.
\newblock \bibinfo{title}{Statistical properties of microcracking in
  polyurethane foams under tensile test, influence of temperature and density}.
\newblock \bibinfo{journal}{Int. J. Frac.} \bibinfo{volume}{140},
  \bibinfo{pages}{87--98}.
\bibitem[{Doussal et~al.(2004)Doussal, Wiese and Chauve}]{LeDoussal}
\bibinfo{author}{Doussal, P.L.}, \bibinfo{author}{Wiese, K.J.},
  \bibinfo{author}{Chauve, P.}, \bibinfo{year}{2004}.
\newblock \bibinfo{title}{Functional renormalization group and the field theory
  of disordered elastic systems}.
\newblock \bibinfo{journal}{Phys. Rev. E} \bibinfo{volume}{68},
  \bibinfo{pages}{026112}.
\bibitem[{Duemmer and Krauth(2005)}]{Duemmer2}
\bibinfo{author}{Duemmer, O.}, \bibinfo{author}{Krauth, W.},
  \bibinfo{year}{2005}.
\newblock \bibinfo{title}{Critical exponents of the driven elastic string in a
  disordered medium}.
\newblock \bibinfo{journal}{Phys. Rev. E} \bibinfo{volume}{71},
  \bibinfo{pages}{061601}.
\bibitem[{Eggers et~al.(2009)Eggers, Indekeu, Meunier and Rolley}]{Eggers}
\bibinfo{author}{Eggers, J.}, \bibinfo{author}{Indekeu, J.},
  \bibinfo{author}{Meunier, J.}, \bibinfo{author}{Rolley, E.},
  \bibinfo{year}{2009}.
\newblock \bibinfo{title}{Wetting and spreading}.
\newblock \bibinfo{journal}{Rev. Mod. Phys.} \bibinfo{volume}{81},
  \bibinfo{pages}{739}.
\bibitem[{Fortin et~al.(2006)Fortin, Stanchits, Dresen and Gu\'eguen}]{Fortin}
\bibinfo{author}{Fortin, J.}, \bibinfo{author}{Stanchits, S.},
  \bibinfo{author}{Dresen, G.}, \bibinfo{author}{Gu\'eguen, Y.},
  \bibinfo{year}{2006}.
\newblock \bibinfo{title}{Acoustic emission and velocities associated with the
  formation of compaction bands in sandstone}.
\newblock \bibinfo{journal}{J. Geophys. Res.} \bibinfo{volume}{111},
  \bibinfo{pages}{B10203}.
\bibitem[{Fortin et~al.(2009)Fortin, Stanchits, Dresen and Gu\'eguen}]{Fortin2}
\bibinfo{author}{Fortin, J.}, \bibinfo{author}{Stanchits, S.},
  \bibinfo{author}{Dresen, G.}, \bibinfo{author}{Gu\'eguen, Y.},
  \bibinfo{year}{2009}.
\newblock \bibinfo{title}{Acoustic emission monitoring during inelastic
  deformation of porous sandstone: Comparison of three modes of deformation}.
\newblock \bibinfo{journal}{Pure Appl. Geophys.} \bibinfo{volume}{166}.
\bibitem[{Fr\'emond and Nedjar(1996)}]{Fremond}
\bibinfo{author}{Fr\'emond, M.}, \bibinfo{author}{Nedjar, B.},
  \bibinfo{year}{1996}.
\newblock \bibinfo{title}{Damage, gradient of damage and principle of virtual
  power}.
\newblock \bibinfo{journal}{Int. J. Solids Struct.} \bibinfo{volume}{33},
  \bibinfo{pages}{1083--1103}.
\bibitem[{Gao and Rice(1989)}]{Gao}
\bibinfo{author}{Gao, H.}, \bibinfo{author}{Rice, J.R.}, \bibinfo{year}{1989}.
\newblock \bibinfo{title}{A first-order perturbation analysis of crack trapping
  by arrays of obstacles}.
\newblock \bibinfo{journal}{J. Appl. Mech.} \bibinfo{volume}{56},
  \bibinfo{pages}{828--836}.
\bibitem[{Garcimartin et~al.(1997)Garcimartin, Guarino, Bellon and
  Ciliberto}]{Garcimartin}
\bibinfo{author}{Garcimartin, A.}, \bibinfo{author}{Guarino, A.},
  \bibinfo{author}{Bellon, L.}, \bibinfo{author}{Ciliberto, .},
  \bibinfo{year}{1997}.
\newblock \bibinfo{title}{Statistical properties of fracture precursors}.
\newblock \bibinfo{journal}{Phys. Rev. Lett.} \bibinfo{volume}{79},
  \bibinfo{pages}{3203--3205}.
\bibitem[{Girard et~al.(2010)Girard, Amitrano and Weiss}]{Girard}
\bibinfo{author}{Girard, L.}, \bibinfo{author}{Amitrano, D.},
  \bibinfo{author}{Weiss, J.}, \bibinfo{year}{2010}.
\newblock \bibinfo{title}{Failure as a critical phenomenon in a progressive
  damage model}.
\newblock \bibinfo{journal}{J. Stat. Mech.} , \bibinfo{pages}{P01013}.
\bibitem[{Godin et~al.(2019)Godin, Reynaud and Fantozzi}]{Godin}
\bibinfo{author}{Godin, N.}, \bibinfo{author}{Reynaud, P.},
  \bibinfo{author}{Fantozzi, G.}, \bibinfo{year}{2019}.
\newblock \bibinfo{title}{Contribution of ae analysis in order to evaluate time
  to failure of ceramic matrix composites}.
\newblock \bibinfo{journal}{Engineering Fracture Mechanics}
  \bibinfo{volume}{210}, \bibinfo{pages}{452--469}.
\bibitem[{Guarino et~al.(2002)Guarino, Ciliberto, Garcimart\'in, Zei and
  Scorretti}]{Guarino2}
\bibinfo{author}{Guarino, A.}, \bibinfo{author}{Ciliberto, S.},
  \bibinfo{author}{Garcimart\'in, A.}, \bibinfo{author}{Zei, M.},
  \bibinfo{author}{Scorretti, M.}, \bibinfo{year}{2002}.
\newblock \bibinfo{title}{Failure time and critical behaviour of fracture
  precursors in heterogeneous materials}.
\newblock \bibinfo{journal}{Eur. Phys. J. B} \bibinfo{volume}{26},
  \bibinfo{pages}{141--151}.
\bibitem[{Guarino et~al.(1998)Guarino, Garcimartin and Ciliberto}]{Guarino}
\bibinfo{author}{Guarino, A.}, \bibinfo{author}{Garcimartin, A.},
  \bibinfo{author}{Ciliberto, S.}, \bibinfo{year}{1998}.
\newblock \bibinfo{title}{An experimental test of the critical behavior of
  fracture precursors}.
\newblock \bibinfo{journal}{Eur. Phys. J. B} \bibinfo{volume}{6},
  \bibinfo{pages}{13--24}.
\bibitem[{Herrmann and Roux(1990)}]{Roux2}
\bibinfo{author}{Herrmann, H.}, \bibinfo{author}{Roux, S.},
  \bibinfo{year}{1990}.
\newblock \bibinfo{title}{Statistical Models for the Fracture of Disordered
  Media}.
\newblock \bibinfo{publisher}{Elsevier}.
\bibitem[{Hirata(1987)}]{hirata1987a}
\bibinfo{author}{Hirata, T.}, \bibinfo{year}{1987}.
\newblock \bibinfo{title}{Omori's power law aftershock sequences of
  microfracturing in rock fracture experiment}.
\newblock \bibinfo{journal}{Journal of Geophysical Research: Solid Earth}
  \bibinfo{volume}{92}, \bibinfo{pages}{6215--6221}.
\bibitem[{Hirata et~al.(1987)Hirata, Satoh and Ito}]{hirata1987b}
\bibinfo{author}{Hirata, T.}, \bibinfo{author}{Satoh, T.},
  \bibinfo{author}{Ito, K.}, \bibinfo{year}{1987}.
\newblock \bibinfo{title}{Fractal structure of spatial distribution of
  microfracturing in rock}.
\newblock \bibinfo{journal}{Geophysical Journal International}
  \bibinfo{volume}{90}, \bibinfo{pages}{369--374}.
\bibitem[{Kachanov(1993)}]{Kachanov4}
\bibinfo{author}{Kachanov, M.}, \bibinfo{year}{1993}.
\newblock \bibinfo{title}{Elastic solids with many cracks and related
  problems}.
\newblock \bibinfo{journal}{Advances in applied mechanics}
  \bibinfo{volume}{30}, \bibinfo{pages}{259--445}.
\bibitem[{Kachanov(2003)}]{Kachanov}
\bibinfo{author}{Kachanov, M.}, \bibinfo{year}{2003}.
\newblock \bibinfo{title}{On the problem of crack interactions and crack
  coalescence}.
\newblock \bibinfo{journal}{Int. J. Frac.} \bibinfo{volume}{120},
  \bibinfo{pages}{537--543}.
\bibitem[{Kandula et~al.(2019)Kandula, Cordonnier, Boller, Weiss, Dysthe and
  Renard}]{kandula2019}
\bibinfo{author}{Kandula, N.}, \bibinfo{author}{Cordonnier, B.},
  \bibinfo{author}{Boller, E.}, \bibinfo{author}{Weiss, J.},
  \bibinfo{author}{Dysthe, D.K.}, \bibinfo{author}{Renard, F.},
  \bibinfo{year}{2019}.
\newblock \bibinfo{title}{Dynamics of microscale precursors during brittle
  compressive failure in carrara marble}.
\newblock \bibinfo{journal}{Journal of Geophysical Research: Solid Earth}
  \bibinfo{volume}{124}, \bibinfo{pages}{6121--6139}.
\bibitem[{Karimi et~al.(2019)Karimi, Amitrano and Weiss}]{Karimi}
\bibinfo{author}{Karimi, K.}, \bibinfo{author}{Amitrano, D.},
  \bibinfo{author}{Weiss, J.}, \bibinfo{year}{2019}.
\newblock \bibinfo{title}{From plastic flow to brittle fracture: Role of
  microscopic friction in amorphous solids}.
\newblock \bibinfo{journal}{Physical Review E} \bibinfo{volume}{100},
  \bibinfo{pages}{012908}.
\bibitem[{Kilburn and Voight(1998)}]{kilburn1998}
\bibinfo{author}{Kilburn, C.R.}, \bibinfo{author}{Voight, B.},
  \bibinfo{year}{1998}.
\newblock \bibinfo{title}{Slow rock fracture as eruption precursor at soufriere
  hills volcano, montserrat}.
\newblock \bibinfo{journal}{Geophysical Research Letters} \bibinfo{volume}{25},
  \bibinfo{pages}{3665--3668}.
\bibitem[{Kun et~al.(2014)Kun, Varga, Lennartz-Sassinek and Main}]{Kun}
\bibinfo{author}{Kun, F.}, \bibinfo{author}{Varga, I.},
  \bibinfo{author}{Lennartz-Sassinek, S.}, \bibinfo{author}{Main, I.G.},
  \bibinfo{year}{2014}.
\newblock \bibinfo{title}{Rupture cascades in a discrete element model of a
  porous sedimentary rock}.
\newblock \bibinfo{journal}{Phys. Rev. Lett.} \bibinfo{volume}{112},
  \bibinfo{pages}{165501}.
\bibitem[{Kun et~al.(2013)Kun, Varga, Lennrtz-Sassinek and Main}]{Kun2}
\bibinfo{author}{Kun, F.}, \bibinfo{author}{Varga, I.},
  \bibinfo{author}{Lennrtz-Sassinek, S.}, \bibinfo{author}{Main, I.G.},
  \bibinfo{year}{2013}.
\newblock \bibinfo{title}{Approach to failure in porous granular materials
  under compression}.
\newblock \bibinfo{journal}{Phys. Rev. E} \bibinfo{volume}{88},
  \bibinfo{pages}{062207}.
\bibitem[{Lawn(1993)}]{Lawn}
\bibinfo{author}{Lawn, B.}, \bibinfo{year}{1993}.
\newblock \bibinfo{title}{Fracture of brittle solids}.
\newblock \bibinfo{publisher}{Cambridge University Press}.
\bibitem[{Le~Priol et~al.(2021)Le~Priol, Le~Doussal and Rosso}]{LePriol}
\bibinfo{author}{Le~Priol, C.}, \bibinfo{author}{Le~Doussal, P.},
  \bibinfo{author}{Rosso, A.}, \bibinfo{year}{2021}.
\newblock \bibinfo{title}{Spatial clustering of depinning avalanches in
  presence of long-range interactions}.
\newblock \bibinfo{journal}{Physical Review Letters} \bibinfo{volume}{126},
  \bibinfo{pages}{025702}.
\bibitem[{Lemaitre(1992)}]{Lemaitre}
\bibinfo{author}{Lemaitre, J.}, \bibinfo{year}{1992}.
\newblock \bibinfo{title}{A course on damage mechanics}.
\newblock Amsterdam, \bibinfo{publisher}{Springer Verlag}.
\bibitem[{Leschhorn et~al.(1997)Leschhorn, Nattermann, Stepanow and
  Tang}]{Leschhorn2}
\bibinfo{author}{Leschhorn, H.}, \bibinfo{author}{Nattermann, T.},
  \bibinfo{author}{Stepanow, S.}, \bibinfo{author}{Tang, L.H.},
  \bibinfo{year}{1997}.
\newblock \bibinfo{title}{Driven interface depinning in a disordered medium}.
\newblock \bibinfo{journal}{Ann. Phys.} \bibinfo{volume}{6},
  \bibinfo{pages}{1--34}.
\bibitem[{Lin et~al.(2014)Lin, Lerner, Rosso and Wyart}]{Lin}
\bibinfo{author}{Lin, J.}, \bibinfo{author}{Lerner, E.},
  \bibinfo{author}{Rosso, A.}, \bibinfo{author}{Wyart, M.},
  \bibinfo{year}{2014}.
\newblock \bibinfo{title}{Scaling description of the yielding transition in
  soft amorphous solids at zero temperature}.
\newblock \bibinfo{journal}{Proc. Nat. Acad. Sci.} \bibinfo{volume}{111},
  \bibinfo{pages}{14382}.
\bibitem[{Lockner(1993)}]{Lockner2}
\bibinfo{author}{Lockner, D.A.}, \bibinfo{year}{1993}.
\newblock \bibinfo{title}{The role of acoustic emission in the study of rock}.
\newblock \bibinfo{journal}{Int. J. Rock Mech. Min. Sci.} \bibinfo{volume}{30},
  \bibinfo{pages}{883--899}.
\bibitem[{Lockner et~al.(1991)Lockner, Byerlee, Kiksenko, Ponomarev and
  Sidorin}]{Lockner}
\bibinfo{author}{Lockner, D.A.}, \bibinfo{author}{Byerlee, J.B.},
  \bibinfo{author}{Kiksenko, V.}, \bibinfo{author}{Ponomarev, A.},
  \bibinfo{author}{Sidorin, A.}, \bibinfo{year}{1991}.
\newblock \bibinfo{title}{Quasi-static fault growth and shear fracture energy
  in granite}.
\newblock \bibinfo{journal}{Nature} \bibinfo{volume}{350},
  \bibinfo{pages}{39--42}.
\bibitem[{Manzato et~al.(2014)Manzato, Alava and Zapperi}]{Manzato}
\bibinfo{author}{Manzato, C.}, \bibinfo{author}{Alava, M.J.},
  \bibinfo{author}{Zapperi, S.}, \bibinfo{year}{2014}.
\newblock \bibinfo{title}{Damage accumulation in quasibrittle fracture}.
\newblock \bibinfo{journal}{Phys. Rev. E} \bibinfo{volume}{90},
  \bibinfo{pages}{012408}.
\bibitem[{Mayya et~al.(2020)Mayya, Berthier and Ponson}]{Mayya}
\bibinfo{author}{Mayya, A.}, \bibinfo{author}{Berthier, E.},
  \bibinfo{author}{Ponson, L.}, \bibinfo{year}{2020}.
\newblock \bibinfo{title}{Proc\'ed\'e et dispositif d’analyse d’une
  structure. french patent application fr2002824}.
\bibitem[{Moreno et~al.(2000)Moreno, G\`{o}mez and Pacheco}]{Moreno}
\bibinfo{author}{Moreno, Y.}, \bibinfo{author}{G\`{o}mez, J.B.},
  \bibinfo{author}{Pacheco, A.F.}, \bibinfo{year}{2000}.
\newblock \bibinfo{title}{Fracture and second-order phase transitions}.
\newblock \bibinfo{journal}{Phys. Rev. Lett.} \bibinfo{volume}{85},
  \bibinfo{pages}{2865}.
\bibitem[{Narayan and Fisher(1993)}]{Narayan}
\bibinfo{author}{Narayan, O.}, \bibinfo{author}{Fisher, D.S.},
  \bibinfo{year}{1993}.
\newblock \bibinfo{title}{Threshold critical dynamics of driven interfaces in
  random media}.
\newblock \bibinfo{journal}{Phys. Rev. B} \bibinfo{volume}{48},
  \bibinfo{pages}{7030--7042}.
\bibitem[{Ojala et~al.(2004)Ojala, Main and Ngwenya}]{ojala2004}
\bibinfo{author}{Ojala, I.O.}, \bibinfo{author}{Main, I.G.},
  \bibinfo{author}{Ngwenya, B.T.}, \bibinfo{year}{2004}.
\newblock \bibinfo{title}{Strain rate and temperature dependence of omori law
  scaling constants of ae data: Implications for earthquake
  foreshock-aftershock sequences}.
\newblock \bibinfo{journal}{Geophysical Research Letters} \bibinfo{volume}{31}.
\bibitem[{Petri et~al.(1994)Petri, Paparao, Vespignani, Alippi and
  Costantini}]{Petri}
\bibinfo{author}{Petri, A.}, \bibinfo{author}{Paparao, G.},
  \bibinfo{author}{Vespignani, A.}, \bibinfo{author}{Alippi, A.},
  \bibinfo{author}{Costantini, M.}, \bibinfo{year}{1994}.
\newblock \bibinfo{title}{Experimental evidence for critical dynamics in
  microfracturing processes}.
\newblock \bibinfo{journal}{Phys. Rev. Lett.} \bibinfo{volume}{73},
  \bibinfo{pages}{3423--2326}.
\bibitem[{Pham et~al.(2011a)Pham, Amor, Marigo and Maurini}]{Pham3}
\bibinfo{author}{Pham, K.}, \bibinfo{author}{Amor, H.},
  \bibinfo{author}{Marigo, J.J.}, \bibinfo{author}{Maurini, C.},
  \bibinfo{year}{2011}a.
\newblock \bibinfo{title}{Gradient damage models and their use to approximate
  brittle fracture}.
\newblock \bibinfo{journal}{International Journal of Damage Mechanics}
  \bibinfo{volume}{20}, \bibinfo{pages}{618--652}.
\bibitem[{Pham et~al.(2011b)Pham, Marigo and Maurini}]{Pham}
\bibinfo{author}{Pham, K.}, \bibinfo{author}{Marigo, J.J.},
  \bibinfo{author}{Maurini, C.}, \bibinfo{year}{2011}b.
\newblock \bibinfo{title}{The issues of the uniqueness and the stability of the
  homogeneous response in the uniaxial tests with gradient damage models}.
\newblock \bibinfo{journal}{J. Mech. Phys. Solids} \bibinfo{volume}{59},
  \bibinfo{pages}{1163--1190}.
\bibitem[{Pijaudier-Cabot and Bazant(1987)}]{Pijaudier}
\bibinfo{author}{Pijaudier-Cabot, G.}, \bibinfo{author}{Bazant, Z.P.},
  \bibinfo{year}{1987}.
\newblock \bibinfo{title}{Nonlocal damage theory}.
\newblock \bibinfo{journal}{J. Eng. Mech.} \bibinfo{volume}{113},
  \bibinfo{pages}{1512--1533}.
\bibitem[{Pijaudier-Cabot and Gr\'egoire(2014)}]{Pijaudier3}
\bibinfo{author}{Pijaudier-Cabot, G.}, \bibinfo{author}{Gr\'egoire, D.},
  \bibinfo{year}{2014}.
\newblock \bibinfo{title}{A review of non local continuum damage: Modeling of
  failure}.
\newblock \bibinfo{journal}{Networks and heterogeneous media}
  \bibinfo{volume}{9}, \bibinfo{pages}{575--597}.
\bibitem[{Planet et~al.(2009)Planet, Santucci and Ortin}]{Planet}
\bibinfo{author}{Planet, R.}, \bibinfo{author}{Santucci, S.},
  \bibinfo{author}{Ortin, J.}, \bibinfo{year}{2009}.
\newblock \bibinfo{title}{Avalanches and non-gaussian fluctuations of the
  global velocity of imbibition fronts}.
\newblock \bibinfo{journal}{Phys. Rev. Lett.} \bibinfo{volume}{102},
  \bibinfo{pages}{094502}.
\bibitem[{Ponson(2016)}]{Ponson19}
\bibinfo{author}{Ponson, L.}, \bibinfo{year}{2016}.
\newblock \bibinfo{title}{Statistical aspects in crack growth phenomena: How
  the fluctuations reveal the failure mechanisms}.
\newblock \bibinfo{journal}{Int. J. Frac.} \bibinfo{volume}{201},
  \bibinfo{pages}{11--27}.
\bibitem[{Ponson and Pindra(2017)}]{Ponson20}
\bibinfo{author}{Ponson, L.}, \bibinfo{author}{Pindra, N.},
  \bibinfo{year}{2017}.
\newblock \bibinfo{title}{Crack propagation through disordered materials as a
  depinning transition: A critical test of the theory}.
\newblock \bibinfo{journal}{Phys. Rev. E} \bibinfo{volume}{(in press)}.
\bibitem[{Pradhan et~al.(2010)Pradhan, Hansen and Chakrabarti}]{Pradhan3}
\bibinfo{author}{Pradhan, S.}, \bibinfo{author}{Hansen, A.},
  \bibinfo{author}{Chakrabarti, B.K.}, \bibinfo{year}{2010}.
\newblock \bibinfo{title}{Failure processes in elastic fiber bundles}.
\newblock \bibinfo{journal}{Rev. Mod. Phys.} \bibinfo{volume}{82},
  \bibinfo{pages}{499--555}.
\bibitem[{Puglisi and Truskinovsky(2005)}]{Puglisi}
\bibinfo{author}{Puglisi, G.}, \bibinfo{author}{Truskinovsky, L.},
  \bibinfo{year}{2005}.
\newblock \bibinfo{title}{Thermodynamics of rate-independent plasticity}.
\newblock \bibinfo{journal}{J. Mech. Phys. Solids} \bibinfo{volume}{53},
  \bibinfo{pages}{655--679}.
\bibitem[{Raischel et~al.(2006)Raischel, Kun and Herrmann}]{Raischel}
\bibinfo{author}{Raischel, F.}, \bibinfo{author}{Kun, F.},
  \bibinfo{author}{Herrmann, H.J.}, \bibinfo{year}{2006}.
\newblock \bibinfo{title}{Failure process of a bundle of plastic fibers}.
\newblock \bibinfo{journal}{Physical Review E} \bibinfo{volume}{73},
  \bibinfo{pages}{066101}.
\bibitem[{Renard et~al.(2018)Renard, Weiss, Mathiesen, Ben-Zion, Kandula and
  Cordonnier}]{Renard2}
\bibinfo{author}{Renard, F.}, \bibinfo{author}{Weiss, J.},
  \bibinfo{author}{Mathiesen, J.}, \bibinfo{author}{Ben-Zion, Y.},
  \bibinfo{author}{Kandula, N.}, \bibinfo{author}{Cordonnier, B.},
  \bibinfo{year}{2018}.
\newblock \bibinfo{title}{Critical evolution of damage toward system-size
  failure in crystalline rock}.
\newblock \bibinfo{journal}{Journal of Geophysical Research: Solid Earth}
  \bibinfo{volume}{123}, \bibinfo{pages}{1969--1986}.
\bibitem[{Rice(1978)}]{Rice8}
\bibinfo{author}{Rice, J.}, \bibinfo{year}{1978}.
\newblock \bibinfo{title}{Thermodynamics of the quasi-static growth of griffith
  cracks}.
\newblock \bibinfo{journal}{J. Mech. Phys. Solids} \bibinfo{volume}{26},
  \bibinfo{pages}{61--78}.
\bibitem[{da~Rocha and Truskinovsky(2020)}]{daRocha}
\bibinfo{author}{da~Rocha, H.B.}, \bibinfo{author}{Truskinovsky, L.},
  \bibinfo{year}{2020}.
\newblock \bibinfo{title}{Rigidity-controlled crossover: from spinodal to
  critical failure}.
\newblock \bibinfo{journal}{Physical review letters} \bibinfo{volume}{124},
  \bibinfo{pages}{015501}.
\bibitem[{Rosso et~al.(2009)Rosso, Doussal and Wiese}]{rosso7}
\bibinfo{author}{Rosso, A.}, \bibinfo{author}{Doussal, P.L.},
  \bibinfo{author}{Wiese, K.J.}, \bibinfo{year}{2009}.
\newblock \bibinfo{title}{Avalanche-size distribution at the depinning
  transition: A numerical test of the theory}.
\newblock \bibinfo{journal}{Phys. Rev. B} \bibinfo{volume}{80},
  \bibinfo{pages}{144204}.
\bibitem[{Rosso et~al.(2003)Rosso, Hartmann and Krauth}]{Rosso9}
\bibinfo{author}{Rosso, A.}, \bibinfo{author}{Hartmann, A.K.},
  \bibinfo{author}{Krauth, W.}, \bibinfo{year}{2003}.
\newblock \bibinfo{title}{Depinning of elastic manifolds}.
\newblock \bibinfo{journal}{Phys. Rev. E} \bibinfo{volume}{67},
  \bibinfo{pages}{021602}.
\bibitem[{Rosti et~al.(2009)Rosti, Illa, Koivisto and Alava}]{Rosti}
\bibinfo{author}{Rosti, J.}, \bibinfo{author}{Illa, X.},
  \bibinfo{author}{Koivisto, J.}, \bibinfo{author}{Alava, M.J.},
  \bibinfo{year}{2009}.
\newblock \bibinfo{title}{Crackling noise and its dynamics in fracture of
  disordered media}.
\newblock \bibinfo{journal}{J. Phys. D: Appl. Phys.} \bibinfo{volume}{42},
  \bibinfo{pages}{214013}.
\bibitem[{Rudnicki and Rice(1975)}]{Rudnicki}
\bibinfo{author}{Rudnicki, J.W.}, \bibinfo{author}{Rice, J.R.},
  \bibinfo{year}{1975}.
\newblock \bibinfo{title}{Conditions for the localization of deformation in
  pressure-sensitive dilatant materials}.
\newblock \bibinfo{journal}{J. Mech. Phys. Solids} \bibinfo{volume}{23},
  \bibinfo{pages}{371--394}.
\bibitem[{Salje et~al.(2017)Salje, Saxena and Planes}]{salje2017}
\bibinfo{author}{Salje, E.K.}, \bibinfo{author}{Saxena, A.},
  \bibinfo{author}{Planes, A.}, \bibinfo{year}{2017}.
\newblock \bibinfo{title}{Avalanches in Functional Materials and Geophysics}.
\newblock \bibinfo{publisher}{Springer}.
\bibitem[{Schmittbuhl et~al.(1995)Schmittbuhl, Roux, Vilotte and
  M{\aa}l{\o}y}]{Schmittbuhl4}
\bibinfo{author}{Schmittbuhl, J.}, \bibinfo{author}{Roux, S.},
  \bibinfo{author}{Vilotte, J.P.}, \bibinfo{author}{M{\aa}l{\o}y, K.J.},
  \bibinfo{year}{1995}.
\newblock \bibinfo{title}{Interfacial crack pinning: effect of nonlocal
  interactions}.
\newblock \bibinfo{journal}{Phys. Rev. Lett.} \bibinfo{volume}{74},
  \bibinfo{pages}{1787--1790}.
\bibitem[{Sethna et~al.(2001)Sethna, Dahmen and Myers}]{Sethna}
\bibinfo{author}{Sethna, J.P.}, \bibinfo{author}{Dahmen, K.A.},
  \bibinfo{author}{Myers, C.R.}, \bibinfo{year}{2001}.
\newblock \bibinfo{title}{Crackling noise}.
\newblock \bibinfo{journal}{Nature} \bibinfo{volume}{410},
  \bibinfo{pages}{241--250}.
\bibitem[{Shekhawat et~al.(2013)Shekhawat, Zapperi and Sethna}]{Shekhawat}
\bibinfo{author}{Shekhawat, A.}, \bibinfo{author}{Zapperi, S.},
  \bibinfo{author}{Sethna, J.P.}, \bibinfo{year}{2013}.
\newblock \bibinfo{title}{From damage percolation to crack nucleation through
  finite size criticality}.
\newblock \bibinfo{journal}{Phys. Rev. Lett.} \bibinfo{volume}{110},
  \bibinfo{pages}{185505}.
\bibitem[{Sornette(2002)}]{Sornette}
\bibinfo{author}{Sornette, D.}, \bibinfo{year}{2002}.
\newblock \bibinfo{title}{Predictability of catastrophic events: Material
  rupture, earthquakes, turbulence, financial crashes, and human birth}.
\newblock \bibinfo{journal}{Proceedings of the National Academy of Sciences}
  \bibinfo{volume}{99}, \bibinfo{pages}{2522--2529}.
\bibitem[{Sornette et~al.(1996)Sornette, Johansen and Bouchaud}]{JPBouchaud2}
\bibinfo{author}{Sornette, D.}, \bibinfo{author}{Johansen, A.},
  \bibinfo{author}{Bouchaud, J.P.}, \bibinfo{year}{1996}.
\newblock \bibinfo{title}{Stock market craches, precursors and replicas}.
\newblock \bibinfo{journal}{J. Phys. I France} \bibinfo{volume}{6},
  \bibinfo{pages}{167--175}.
\bibitem[{Tal et~al.(2016)Tal, Evans and Mok}]{Tal}
\bibinfo{author}{Tal, Y.}, \bibinfo{author}{Evans, B.}, \bibinfo{author}{Mok,
  U.}, \bibinfo{year}{2016}.
\newblock \bibinfo{title}{Direct observations of damage during unconfined
  brittle failure of carrara marble}.
\newblock \bibinfo{journal}{J. Geophys. Res. Solid Earth}
  \bibinfo{volume}{121}, \bibinfo{pages}{1584--1609}.
\bibitem[{Tang(1997)}]{Tang}
\bibinfo{author}{Tang, C.A.}, \bibinfo{year}{1997}.
\newblock \bibinfo{title}{Numerical simulation of progressive rock failure and
  associated seismicity}.
\newblock \bibinfo{journal}{Int. J. Rock Mech. Min. Sci.} \bibinfo{volume}{34},
  \bibinfo{pages}{249--261}.
\bibitem[{Thilakarathna et~al.(2020)Thilakarathna, Baduge, Mendis, Vimonsatit
  and Lee}]{Thilakarathna}
\bibinfo{author}{Thilakarathna, P.}, \bibinfo{author}{Baduge, K.K.},
  \bibinfo{author}{Mendis, P.}, \bibinfo{author}{Vimonsatit, V.},
  \bibinfo{author}{Lee, H.}, \bibinfo{year}{2020}.
\newblock \bibinfo{title}{Mesoscale modelling of concrete--a review of geometry
  generation, placing algorithms, constitutive relations and applications}.
\newblock \bibinfo{journal}{Engineering Fracture Mechanics}
  \bibinfo{volume}{231}, \bibinfo{pages}{106974}.
\bibitem[{Vandembroucq and Roux(2011)}]{Vandembroucq4}
\bibinfo{author}{Vandembroucq, D.}, \bibinfo{author}{Roux, S.},
  \bibinfo{year}{2011}.
\newblock \bibinfo{title}{Mechanical noise dependent aging and shear banding
  behavior of a mesoscopic model of amorphous plasticity}.
\newblock \bibinfo{journal}{Phys. Rev. B} \bibinfo{volume}{84},
  \bibinfo{pages}{134210}.
\bibitem[{Voight(1988)}]{voight1988}
\bibinfo{author}{Voight, B.}, \bibinfo{year}{1988}.
\newblock \bibinfo{title}{A method for prediction of volcanic eruptions}.
\newblock \bibinfo{journal}{Nature} \bibinfo{volume}{332},
  \bibinfo{pages}{125--130}.
\bibitem[{Vu et~al.(2019)Vu, Amitrano, Pl\'{e} and Weiss}]{Vu}
\bibinfo{author}{Vu, C.C.}, \bibinfo{author}{Amitrano, D.},
  \bibinfo{author}{Pl\'{e}, O.}, \bibinfo{author}{Weiss, J.},
  \bibinfo{year}{2019}.
\newblock \bibinfo{title}{Compressive failure as a critical transition:
  Experimental evidence and mapping onto the universality class of depinning}.
\newblock \bibinfo{journal}{Phys. Rev. Lett.} \bibinfo{volume}{122},
  \bibinfo{pages}{015502}.
\bibitem[{Weiss et~al.(2014)Weiss, Girard, Gimbert, Amitrano and
  Vandembroucq}]{Weiss2}
\bibinfo{author}{Weiss, J.}, \bibinfo{author}{Girard, L.},
  \bibinfo{author}{Gimbert, F.}, \bibinfo{author}{Amitrano, D.},
  \bibinfo{author}{Vandembroucq, D.}, \bibinfo{year}{2014}.
\newblock \bibinfo{title}{(finite) statistical size effects on compressive
  strength}.
\newblock \bibinfo{journal}{Proc. Nat. Acad. Sci.} \bibinfo{volume}{111},
  \bibinfo{pages}{6231--6236}.
\bibitem[{Wiese(2021)}]{Wiese2}
\bibinfo{author}{Wiese, K.J.}, \bibinfo{year}{2021}.
\newblock \bibinfo{title}{Theory and experiments for disordered elastic
  manifolds, depinning, avalanches, and sandpiles}.
\newblock \bibinfo{journal}{arXiv preprint arXiv:2102.01215} .
\bibitem[{Zapperi et~al.(1998)Zapperi, Cizeau, Durin and Stanley}]{Zapperi2}
\bibinfo{author}{Zapperi, S.}, \bibinfo{author}{Cizeau, P.},
  \bibinfo{author}{Durin, G.}, \bibinfo{author}{Stanley, H.E.},
  \bibinfo{year}{1998}.
\newblock \bibinfo{title}{Dynamics of ferromagnetic domain wall: avalanches,
  depinning transition and the barkhausen effect}.
\newblock \bibinfo{journal}{Phys. Rev. E} \bibinfo{volume}{58},
  \bibinfo{pages}{6353--6366}.
\bibitem[{Zapperi et~al.(1997a)Zapperi, Ray, Stanley and Vespignani}]{Zapperi5}
\bibinfo{author}{Zapperi, S.}, \bibinfo{author}{Ray, P.},
  \bibinfo{author}{Stanley, H.E.}, \bibinfo{author}{Vespignani, A.},
  \bibinfo{year}{1997}a.
\newblock \bibinfo{title}{First-order transition in the breakdown of disordered
  media}.
\newblock \bibinfo{journal}{Phys. Rev. Lett.} \bibinfo{volume}{78},
  \bibinfo{pages}{1408}.
\bibitem[{Zapperi et~al.(1997b)Zapperi, Vespignani and Stanley}]{Zapperi3}
\bibinfo{author}{Zapperi, S.}, \bibinfo{author}{Vespignani, A.},
  \bibinfo{author}{Stanley, H.E.}, \bibinfo{year}{1997}b.
\newblock \bibinfo{title}{Plasticity and avalanches behavior in microfracturing
  phenomena}.
\newblock \bibinfo{journal}{Nature} \bibinfo{volume}{388},
  \bibinfo{pages}{658--660}.

\end{thebibliography}

\end{document}